\documentclass[twocolumn]{aastex631}
\pdfoutput=1
\usepackage{amsmath}
\usepackage{amsfonts}
\usepackage{amssymb}
\usepackage{chngcntr}

\usepackage{graphicx}
\usepackage[varg]{txfonts}
\usepackage{mwe} 
\usepackage{url}
\usepackage{longtable}
\usepackage{float}

\usepackage{makecell}
\usepackage{xspace}
\usepackage{xcolor}

\newcommand{\Q}{\ensuremath{Q_{\mathrm{ion}}}\xspace}
\newcommand{\fedd}{\ensuremath{f_{\mathrm{Edd}}}\xspace}
\newcommand{\Mdot}{\ensuremath{\dot{M}}\xspace}
\newcommand{\A}{\ensuremath{\mathrm{\AA}}\xspace}
\newcommand{\rin}{\ensuremath{r_{\mathrm{in}}}\xspace}
\newcommand{\rout}{\ensuremath{r_{\mathrm{out}}}\xspace}
\newcommand{\Nion}{\ensuremath{\dot{N}_{\mathrm{ion}}}\xspace}
\newcommand{\MMedd}{\ensuremath{\Mdot/\Mdot_{\rm{Edd}}}\xspace}
\newcommand{\Nunits}{\ensuremath{\mathrm{s}^{-1}\mathrm{Mpc}^{-3}}\xspace}
\newcommand{\Rion}{\ensuremath{R_{\mathrm{ion}}}\xspace}

\newcommand{\fesc}{\ensuremath{f_{\mathrm{esc}}}\xspace}
\newcommand{\xiion}{\ensuremath{\xi_{\mathrm{ion}}}\xspace}

\newcommand{\todo}{\ifmmode \text{\color{red}{\Huge(\bullet)}} \else {\color{red}{\Huge$\bullet$}}\fi}
\newcommand{\tido}{\ifmmode {{\color{red}\bullet}} \else \color{red}{$\bullet$}\fi}

\newcommand{\E        }[1]{\ifmmode 10^{#1} \else $10^{#1}$\fi}
\newcommand{\tE        }[1]{\ifmmode \times10^{#1} \else $\times10^{#1}$\fi}
\newcommand{\til}{\ifmmode \sim \else $\sim$\fi}
\renewcommand{\~} {\ifmmode \sim \else $\sim$\fi}

\newcommand{\pc}	{\ifmmode {\rm pc} \else pc\fi}
\newcommand{\kpc}	{\ifmmode {\rm kpc} \else kpc\fi}
\newcommand{\ld}	{\ifmmode {\rm l.d.} \else l.d.\fi}
\newcommand{\kms}	{\ifmmode {\rm km\,s}^{-1} \else km\,s$^{-1}$\fi}
\newcommand{\cc}	{\ifmmode {\rm cm}^{-3}    \else cm$^{-3}$\fi}
\newcommand{\cmii}	{\ifmmode {\rm cm}^{-2}    \else cm$^{-2}$\fi}
\newcommand{\ergs}	{\ifmmode {\rm erg\,s}^{-1} \else erg s$^{-1}$\fi}
\newcommand{\ergcms}	{\ifmmode {\rm erg\,cm}^{-2}\,{\rm s}^{-1} \else erg\,cm$^{-2}$\,s$^{-1}$\fi}
\newcommand{\ergcmsA}	{\ifmmode {\rm erg\,cm}^{-2}\,{\rm s}^{-1}\,{\rm\AA}^{-1}
\else erg\,cm$^{-2}$\,s$^{-1}$\,\AA$^{-1}$\fi}
\newcommand{  \ergcmsHz  }{\ifmmode{\rm erg\,cm}^{-2}\,{\rm s}^{-1}\,{\rm Hz}^{-1}
                       \else ergs\,cm$^{-2}$\,s$^{-1}$\,Hz$^{-1}$\fi}
\newcommand{\kev}	{\ifmmode {\rm keV} \else keV\fi}

\newcommand{\mic}	{\ifmmode {\rm \mu m} \else $\mu$m\fi}
\newcommand{\vFWHM}	{\ifmmode v_{\mbox{\tiny FWHM}} \else $v_{\mbox{\tiny FWHM}}$\fi}
\newcommand{\vBLR}	{\ifmmode v_{\mbox{\tiny BLR}} \else $v_{\mbox{\tiny BLR}}$\fi}
\newcommand{\sigBLR}	{\ifmmode \sigma_{\mbox{\tiny BLR}} \else $\sigma_{\mbox{\tiny BLR}}$\fi}
\newcommand{\vNLR}	{\ifmmode v_{\mbox{\tiny NLR}} \else $v_{\mbox{\tiny NLR}}$\fi}
\newcommand{\tauBLR}	{\ifmmode \tau_{\mbox{\tiny BLR}} \else $\tau_{\mbox{\tiny BLR}}$\fi}

\newcommand{\Hubble}	{\ifmmode {\rm km\,s}^{-1}\,{\rm Mpc}^{-1} \else km\,s$^{-1}$\,Mpc$^{-1}$\fi}
\newcommand{\NDunit}	{\ifmmode {\rm Mpc}^{-3} \else Mpc$^{-3}$\fi}
\newcommand{\LFunit}	{\ifmmode {\rm Mpc}^{-3}\,{\rm mag}^{-1} \else Mpc$^{-3}$\,mag$^{-1}$\fi}
\newcommand{\MFunit}	{\ifmmode {\rm Mpc}^{-3}\,{\rm dex}^{-1} \else Mpc$^{-3}$\,dex$^{-1}$\fi}

\newcommand{\Msun}{\ifmmode M_{\odot} \else $M_{\odot}$\fi}
\newcommand{\Lsun}{\ifmmode L_{\odot} \else $L_{\odot}$\fi}
\newcommand{\Zsun}{\ifmmode Z_{\odot} \else $Z_{\odot}$\fi}
\newcommand{\mpyr}{\ifmmode \Msun\,{\rm yr}^{-1} \else $\Msun\,{\rm yr}^{-1}$\fi}

\newcommand{\qnote}{\ifmmode q_{0} \else $q_{0}$\fi}
\newcommand{\Hnote}{\ifmmode H_{0} \else $H_{0}$\fi}
\newcommand{\hnote}{\ifmmode h_{0} \else $h_{0}$\fi}
\newcommand{\anote}{\ifmmode a_{0} \else $a_{0}$\fi}
\newcommand{\tnote}{\ifmmode t_{0} \else $t_{0}$\fi}

\newcommand{  \Halpha   }{\ifmmode {\rm H}\alpha \else H$\alpha$\fi}

\newcommand{  \ha       }{\Halpha}
\newcommand{  \Hbeta    }{\ifmmode {\rm H}\beta \else H$\beta$\fi}

\newcommand{  \hb       }{\Hbeta}
\newcommand{  \Hgamma   }{\ifmmode {\rm H}\gamma \else H$\gamma$\fi}
\newcommand{  \Hdelta   }{\ifmmode {\rm H}\delta \else H$\delta$\fi}
\newcommand{  \Lya      }{\ifmmode {\rm Ly}\alpha \else Ly$\alpha$\fi}
\newcommand{  \Lyb      }{\ifmmode {\rm Ly}\beta \else Ly$\beta$\fi}
\newcommand{  \Pa       }{\ifmmode {\rm P}\alpha \else P$\alpha$\fi}
\newcommand{  \Pb       }{\ifmmode {\rm P}\beta \else P$\beta$\fi}
\newcommand{  \Bra      }{\ifmmode {\rm Br}\alpha \else Br$\alpha$\fi}
\newcommand{  \Brg      }{\ifmmode {\rm Br}\gamma \else Br$\gamma$\fi}
\newcommand{  \hii      }{\ifmmode {\rm H}\,\textsc{ii} \else H\,\textsc{ii}\fi}
\newcommand{  \hei      }{\ifmmode {\rm He}\,\textsc{i} \else He\,\textsc{i}\fi}
\newcommand{  \heii     }{\ifmmode {\rm He}\,\textsc{ii} \else He\,\textsc{ii}\fi}
\newcommand{  \HeIIuv   }{\ifmmode {\rm He}\,\textsc{ii}\,\lambda1640 \else He\,\textsc{ii}\,$\lambda1640$\fi}
\newcommand{  \HeIIop   }{\ifmmode {\rm He}\,\textsc{ii}\,\lambda4686 \else He\,\textsc{ii}\,$\lambda4686$\fi}
\newcommand{  \CII	}{\ifmmode \left[{\rm C}\,\textsc{ii}\right]\,\lambda157.74\,\mu{\rm m} \else [C\,{\sc ii}]\ $\lambda157.74\,\mu{\rm m}$\fi}
\newcommand{  \cii	}{\ifmmode \left[{\rm C}\,\textsc{ii}\right] \else [C\,{\sc ii}]\fi}

\newcommand{  \ciii     }{\ifmmode {\rm C}\,\textsc{iii}\right] \else C\,\textsc{iii}]\fi}
\newcommand{  \CIII     }{\ifmmode {\rm C}\,\textsc{iii}\right]\,\lambda1909 \else C\,\textsc{iii}]\,$\lambda1909$\fi}
\newcommand{  \civ      }{\ifmmode {\rm C}\,\textsc{iv}  \else C\,\textsc{iv}\fi}
\newcommand{  \CIV      }{\ifmmode {\rm C}\,\textsc{iv}\,\lambda1549 \else C\,\textsc{iv}\,$\lambda1549$\fi}
\newcommand{  \NIIopt   }{\ifmmode \left[{\rm N}\,\textsc{ii}\right]\,\lambda6584 \else [N\,\textsc{ii}]\,$\lambda6584$\fi}
\newcommand{  \nii      }{\ifmmode \left[{\rm N}\,\textsc{ii}\right]  \else [N\,\textsc{ii}]\fi}
\newcommand{  \niii     }{\ifmmode {\rm N}\,\textsc{iii} \else N\,\textsc{iii}\fi}
\newcommand{  \NIII     }{\ifmmode {\rm N}\,\textsc{iii}\,\lambda4640 \else N\,\textsc{iii}\,$\lambda4640$\fi}
\newcommand{  \niv      }{\ifmmode {\rm N}\,\textsc{iv}  \else N\,\textsc{iv}\fi}
\newcommand{  \NIVuv    }{\ifmmode {\rm N}\,\textsc{iv}\,\lambda1486 \else N\,\textsc{iv}\,$\lambda1486$\fi}
\newcommand{  \nv       }{\ifmmode {\rm N}\,\textsc{v}   \else N\,\textsc{v}\fi}
\newcommand{\oi}{\ifmmode \left[{\rm O}\,\textsc{i}\right] \else [O\,{\sc i}]\fi}
\newcommand{\OI}{\ifmmode \left[{\rm O}\,\textsc{i}\right]\,\lambda6300 \else [O\,{\sc i}]$\,\lambda6300$\fi}
\newcommand{\oii}{\ifmmode \left[{\rm O}\,\textsc{ii}\right] \else [O\,{\sc ii}]\fi}
\newcommand{\OII}{\ifmmode \left[{\rm O}\,\textsc{ii}\right]\,\lambda3727 \else [O\,{\sc ii}]\,$\lambda3727$\fi}
\newcommand{\oiii}{\ifmmode \left[{\rm O}\,\textsc{iii}\right] \else [O\,{\sc iii}]\fi}
\newcommand{\OIII}{\ifmmode \left[{\rm O}\,\textsc{iii}\right]\,\lambda5007 \else [O\,{\sc iii}]\,$\lambda5007$\fi}
\newcommand{  \OIIIbf   }{\ifmmode {\rm O}\,\textsc{iii}\,\lambda3133 \else O\,\textsc{iii}\,$\lambda3133$\fi}
\newcommand{  \OIIIuv   }{\ifmmode {\rm O}\,\textsc{iii}\,\lambda1663 \else O\,\textsc{iii}\,$\lambda1663$\fi}
\newcommand{  \oiv      }{\ifmmode {\rm O}\,\textsc{iv}  \else O\,\textsc{iv}\fi}
\newcommand{  \OIVuv    }{\ifmmode {\rm O}\,\textsc{iv}\,\lambda1402  \else O\,\textsc{iv}\,$\lambda1402$\fi}
\newcommand{  \OIVIR    }{\ifmmode {\rm O}\,\textsc{iv}\,25.9\,\mu {\rm m} \else O\,\textsc{iv}\,$25.9\,\mu$m\fi}
\newcommand{  \ovi      }{\ifmmode {\rm O}\,\textsc{vi}   \else O\,\textsc{vi}\fi}
\newcommand{  \Ovi      }{\ifmmode {\rm O}\,\textsc{vi}\,\lambda1035 \else O\,\textsc{vi}\,$\lambda1035$\fi}
\newcommand{  \nei      }{\ifmmode {\rm Ne}\,\textsc{i}   \else Ne\,\textsc{i}\fi}
\newcommand{  \neii     }{\ifmmode {\rm Ne}\,\textsc{ii}  \else Ne\,\textsc{ii}\fi}
\newcommand{  \NeiiIR   }{\ifmmode {\rm Ne}\,\textsc{ii}\,12.8\,\mu {\rm m} \else Ne\,\textsc{ii}\,$12.8\,\mu$m\fi}
\newcommand{  \neiii    }{\ifmmode {\rm Ne}\,\textsc{iii} \else Ne\,\textsc{iii}\fi}
\newcommand{  \neiv     }{\ifmmode {\rm Ne}\,\textsc{iv}  \else Ne\,\textsc{iv}\fi}
\newcommand{  \nev      }{\ifmmode {\rm Ne}\,\textsc{v}   \else Ne\,\textsc{v}\fi}
\newcommand{  \NevIR    }{\ifmmode {\rm Ne}\,\textsc{v}\,24.3\,\mu {\rm m} \else Ne\,\textsc{v}\,$24.3\,\mu$m\fi}
\newcommand{  \nevi     }{\ifmmode {\rm Ne}\,\textsc{vi}  \else Ne\,\textsc{vi}\fi}
\newcommand{  \mgi      }{\ifmmode {\rm Mg}\,\textsc{i} \else Mg\,\textsc{i}\fi}
\newcommand{  \mgii     }{\ifmmode {\rm Mg}\,\textsc{ii} \else Mg\,\textsc{ii}\fi}
\newcommand{  \MgII     }{\ifmmode {\rm Mg}\,\textsc{ii}\,\lambda2798 \else Mg\,\textsc{ii}\,$\lambda2798$\fi}
\newcommand{  \sii      }{\ifmmode {\rm S}\,\textsc{ii} \else S\,\textsc{ii}\fi}
\newcommand{  \siii     }{\ifmmode {\rm S}\,\textsc{iii} \else S\,\textsc{iii}\fi}
\newcommand{  \siv      }{\ifmmode {\rm S}\,\textsc{iv} \else S\,\textsc{iv}\fi}
\newcommand{  \sili     }{\ifmmode {\rm Si}\,\textsc{i}   \else Si\,\textsc{i}\fi}
\newcommand{  \silii    }{\ifmmode {\rm Si}\,\textsc{ii}  \else Si\,\textsc{ii}\fi}
\newcommand{  \Siliv    }{\ifmmode {\rm Si}\,\textsc{iv}  \else Si\,\textsc{iv}\fi}
\newcommand{  \SilIVuv  }{\ifmmode {\rm Si}\,\textsc{iv}\,\lambda1400  \else Si\,\textsc{iv}\,$\lambda1400$\fi}
\newcommand{  \AlIII   }{\ifmmode {\rm Al}\,\textsc{iii}\,\lambda1857 \else Al\,\textsc{iii}\,$\lambda1857$\fi}
\newcommand{  \Aliii   }{\ifmmode {\rm Al}\,\textsc{iii} \else Al\,\textsc{iii}\fi}
\newcommand{  \caii     }{\ifmmode {\rm Ca}\,\textsc{ii} \else Ca\,\textsc{ii}\fi}
\newcommand{  \feii     }{\ifmmode {\rm Fe}\,\textsc{ii} \else Fe\,\textsc{ii}\fi}
\newcommand{  \feiii    }{\ifmmode {\rm Fe}\,\textsc{iii} \else Fe\,\textsc{iii}\fi}
\newcommand{  \Kalpha   }{\ifmmode {\rm K}\alpha \else K$\alpha$\fi}

\newcommand{ \Lhb   }{\ifmmode L_{\hb} \else $L_{\hb}$\fi}
\newcommand{ \Lha   }{\ifmmode L_{\ha} \else $L_{\ha}$\fi}
\newcommand{ \fwhb  }{\ifmmode {\rm FWHM}\left(\hb\right) \else FWHM(\hb)\fi}
\newcommand{\sighb  }{\ifmmode \sigma\left(\hb\right) \else $\sigma\left(\hb\right)$\fi}
\newcommand{ \ewhb  }{\ifmmode {\rm EW}\left(\hb\right) \else EW(\hb)\fi}
\newcommand{ \fwha  }{\ifmmode {\rm FWHM}\left(\ha\right) \else FWHM(\ha)\fi}
\newcommand{ \ewha  }{\ifmmode {\rm EW}\left(\ha\right) \else EW(\ha)\fi}
\newcommand{ \Lmg   }{\ifmmode L\left(\mgii\right) \else $L\left(\mgii\right)$\fi}
\newcommand{ \fwmg  }{\ifmmode {\rm FWHM}\left(\mgii\right) \else FWHM(\mgii)\fi}
\newcommand{ \Lciv  }{\ifmmode L\left(\civ\right) \else $L\left(\civ\right)$\fi}
\newcommand{ \fwciv }{\ifmmode {\rm FWHM}\left(\civ\right) \else FWHM(\civ)\fi}
\newcommand{ \fwhm  }{\ifmmode {\rm FWHM} \else FWHM\fi} 
\newcommand{ \voff  }{\ifmmode v_{\rm off} \else $v_{\rm off}$\fi} 
\newcommand{ \vmax  }{\ifmmode v_{\rm max} \else $v_{\rm max}$\fi} 

\newcommand{ \mumg  }{\ifmmode \mu\left(\mgii\right) \else $\mu\left(\mgii\right)$\fi}
\newcommand{ \fmg   }{\ifmmode f\left(\mgii\right) \else $f\left(\mgii\right)$\fi}
\newcommand{ \muciv }{\ifmmode \mu\left(\civ\right) \else $\mu\left(\civ\right)$\fi}
\newcommand{ \fciv  }{\ifmmode f\left(\civ\right) \else $f\left(\civ\right)$\fi}
\newcommand{  \auvo     }{\ifmmode \alpha_{\nu,{\rm UVO}} \else $\alpha_{\nu,{\rm UVO}}$\fi}
\newcommand{  \Ledd     }{\ifmmode L_{\rm Edd} \else $L_{\rm Edd}$\fi}
\newcommand{  \lamLlam  }{\ifmmode \lambda L_{\lambda} \else $\lambda L_{\lambda}$\fi}
\newcommand{  \lLl      }{\ifmmode \lambda L_{\lambda} \else $\lambda L_{\lambda}$\fi}
\newcommand{  \nuLnu    }{\ifmmode \nu L_{\nu} \else $\nu L_{\nu}$\fi}
\newcommand{  \nLn      }{\ifmmode \nu L_{\nu} \else $\nu L_{\nu}$\fi}
\newcommand{  \Luv      }{\ifmmode L_{1450} \else $L_{1450}$\fi}
\newcommand{  \Lop      }{\ifmmode L_{5100} \else $L_{5100}$\fi}
\newcommand{  \lLop     }{\ifmmode \log\left(\Lop/\ergs\right) \else $\log\left(\Lop/\ergs\right)$\fi}
\newcommand{  \Lthree   }{\ifmmode L_{3000} \else $L_{3000}$\fi}
\newcommand{  \lLthree  }{\ifmmode \log\left(\Lthree/\ergs\right) \else $\log\left(\Lthree/\ergs\right)$\fi}
\newcommand{  \Lsix      }{\ifmmode L_{6200} \else $L_{6200}$\fi}
\newcommand{  \lLisx     }{\ifmmode \log\left(\Lop/\ergs\right) \else $\log\left(\Lop/\ergs\right)$\fi}
\newcommand{  \Lxray    }{\ifmmode L_{\rm X} \else $L_{\rm X}$\fi}

\newcommand{  \Lhard    }{\ifmmode L_{\rm 2-10} \else $L_{\rm 2-10}$\fi}
\newcommand{  \Lsoft    }{\ifmmode L_{\rm 0.5-2} \else $L_{\rm 0.5-2}$\fi}

\newcommand{\Fthree}{\ifmmode F_{3000} \else $F_{3000}$\fi}
\newcommand{\fuv}{\ifmmode f_{\lambda}\left(1450{\rm \AA}\right) \else $f_{\lambda}\left(1450 {\rm \AA}\right)$\fi}
\newcommand{\fthree}{\ifmmode f_{\lambda}\left(3000{\rm \AA}\right) \else $f_{\lambda}\left(3000{\rm \AA}\right)$\fi}
\newcommand{\fH}{\ifmmode f_{\lambda}\left(1.65\micron\right) \else
$f_{\lambda}\left(1.65\micron\right)$\fi}

\newcommand{\fbol}{\ifmmode f_{\rm bol} \else $f_{\rm bol}$\fi}
\newcommand{\fbolwv}{\ifmmode f_{\rm bol}\left(\lambda\right) \else $f_{\rm bol}\left(\lambda\right)$\fi}
\newcommand{\fbolopt}{\ifmmode f_{\rm bol}\left(5100{\rm \AA}\right) \else $f_{\rm bol}\left(5100{\rm \AA}\right)$\fi}
\newcommand{\fbolthree}{\ifmmode f_{\rm bol}\left(3000{\rm \AA}\right) \else $f_{\rm bol}\left(3000{\rm \AA}\right)$\fi}
\newcommand{\fboluv}{\ifmmode f_{\rm bol}\left(1450{\rm \AA}\right) \else $f_{\rm bol}\left(1450{\rm \AA}\right)$\fi}

\newcommand{\fbolbat}{\ifmmode f_{\rm bol}\left(14-150\,\kev\right) \else $f_{\rm bol}\left(14-150\,\kev\right)$\fi}
\newcommand{\fbolhard}{\ifmmode f_{\rm bol}\left(2-10\,\kev\right) \else $f_{\rm bol}\left(2-10\,\kev\right)$\fi}

\newcommand{\fobs}{\ifmmode f_{\rm obs} \else $f_{\rm obs}$\fi}

\newcommand{  \mbh      }{\ifmmode M_{\rm BH} \else $M_{\rm BH}$\fi}
\newcommand{  \lmbh     }{\ifmmode \log\left(\mbh/\Msun\right) \else $\log\left(\mbh/\Msun\right)$\fi} 
\newcommand{  \lledd    }{\ifmmode L/L_{\rm Edd} \else $L/L_{\rm Edd}$\fi}
\newcommand{  \mmedd    }{\ifmmode \dot{m}/\dot{m}_{\rm \,Edd} \else $\dot{m}/\dot{m}_{\rm \,Edd}$\fi}
\newcommand{  \Lbol     }{\ifmmode L_{\rm bol} \else $L_{\rm bol}$\fi}
\newcommand{  \lbol     }{\ifmmode L_{\rm bol} \else $L_{\rm bol}$\fi}
\newcommand{  \lLbol    }{\ifmmode \log\left(\Lbol/\ergs\right) \else $\log\left(\Lbol/\ergs\right)$\fi} 
\newcommand{  \Lagn     }{\ifmmode L_{\rm AGN} \else $L_{\rm AGN}$\fi}
\newcommand{  \lagn     }{\ifmmode L_{\rm AGN} \else $L_{\rm AGN}$\fi}

\newcommand{  \tgrow     }{\ifmmode t_{\rm growth} \else $t_{\rm growth}$\fi}
\newcommand{  \tAD     }{\ifmmode t_{\rm acc} \else $t_{\rm acc}$\fi}
\newcommand{  \tacc    }{\ifmmode t_{\rm acc} \else $t_{\rm acc}$\fi}
\newcommand{  \tUni      }{\ifmmode t_{\rm Universe} \else $t_{\rm Universe}$\fi}

\newcommand{  \Mdotin	}{\ifmmode \dot{M}_{\rm infall} \else $\dot{M}_{\rm infall}$\fi}
\newcommand{  \Mdotbh	}{\ifmmode \dot{M}_{\rm BH} \else $\dot{M}_{\rm BH}$\fi}
\newcommand{  \Mdotad	}{\ifmmode \dot{M}_{\rm AD} \else $\dot{M}_{\rm AD}$\fi}
\newcommand{  \Mdotacc	}{\ifmmode \dot{M}_{\rm acc} \else $\dot{M}_{\rm acc}$\fi}
\newcommand{  \Mdotthin	}{\ifmmode \dot{M}_{\rm thin} \else $\dot{M}_{\rm thin}$\fi}
\newcommand{  \Mdotdisk	}{\ifmmode \dot{M}_{\rm disk} \else $\dot{M}_{\rm disk}$\fi}

\newcommand{  \Mindot	}{\ifmmode \dot{M}_{\rm infall} \else $\dot{M}_{\rm infall}$\fi}
\newcommand{  \Mbhdot	}{\ifmmode \dot{M}_{\rm BH} \else $\dot{M}_{\rm BH}$\fi}
\newcommand{  \Maddot	}{\ifmmode \dot{M}_{\rm AD} \else $\dot{M}_{\rm AD}$\fi}
\newcommand{  \Maccdot	}{\ifmmode \dot{M}_{\rm acc} \else $\dot{M}_{\rm acc}$\fi}
\newcommand{  \Mthdot	}{\ifmmode \dot{M}_{\rm thin} \else $\dot{M}_{\rm thin}$\fi}
\newcommand{  \Mdsdot	}{\ifmmode \dot{M}_{\rm disk} \else $\dot{M}_{\rm disk}$\fi}

\newcommand{  \as	}{\ifmmode a_{\rm *} \else $a_{\rm *}$\fi}
\newcommand{  \avec	}{\ifmmode \vec{a}_{\rm *} \else $\vec{a}_{\rm *}$\fi}
\newcommand{  \re	}{\ifmmode \eta      	 \else $\eta$\fi}
\newcommand{  \RISCO	}{\ifmmode R_{\rm ISCO}  \else $R_{\rm ISCO}$\fi}

\newcommand{  \mseed    }{\ifmmode M_{\rm seed} \else $M_{\rm seed}$\fi}
\newcommand{  \mbul     }{\ifmmode M_{\rm bulge} \else $M_{\rm bulge}$\fi} 
\newcommand{  \mstar    }{\ifmmode M_{*} \else $M_{*}$\fi} 
\newcommand{  \mgal     }{\ifmmode M_{*} \else $M_{*}$\fi} 
\newcommand{  \mhost    }{\ifmmode M_{\rm host} \else $M_{\rm host}$\fi}
\newcommand{  \mmsmall  }{\ifmmode M_{\rm BH}/M_{*} \else $M_{\rm BH}/M_{*}$\fi}
\newcommand{  \mmlarge  }{\ifmmode M_{*}/M_{\rm BH} \else $M_{*}/M_{\rm BH}$\fi}

\newcommand{  \mmdotlarge}{\ifmmode \dot{M}_*/\Mbhdot \else $\dot{M}_*/\Mbhdot$\fi}
\newcommand{  \mmdotsmall}{\ifmmode \Mbhdot/\dot{M}_* \else $\Mbhdot/\dot{M}_*$\fi}

\newcommand{  \mmwp     }{\ifmmode \left(M_{*}/M_{\rm BH}\right) \else $\left(M_{*}/M_{\rm BH}\right)$\fi}
\newcommand{  \ml       }{\ifmmode M_{*}/L_{*} \else $M_{*}/L_{*}$\fi}
\newcommand{  \mlwp     }{\ifmmode \left(M_{*}/L\right) \else $\left(M_{*}/L\right)$\fi}
\newcommand{  \mlk      }{\ifmmode \left(M_{*}/L_{K}\right) \else $\left(M_{*}/L_{K}\right)$\fi}
\newcommand{  \sigs     }{\ifmmode \sigma_{*} \else $\sigma_{*}$\fi}
\newcommand{  \Reff     }{\ifmmode R_{\rm e} \else $R_{\rm e}$\fi}
\newcommand{  \Rvir     }{\ifmmode R_{\rm vir} \else $R_{\rm vir}$\fi}
\newcommand{  \Rtwo     }{\ifmmode R_{200} \else $R_{200}$\fi}
\newcommand{  \Rfive    }{\ifmmode R_{500} \else $R_{500}$\fi}
\newcommand{  \Rgrp     }{\ifmmode R_{\rm grp} \else $R_{\rm grp}$\fi}
\newcommand{  \nser     }{\ifmmode n_{\rm s} \else $n_{\rm s}$\fi}
\newcommand{  \LSF      }{\ifmmode L_{\rm SF}  \else $L_{\rm SF}$\fi}
\newcommand{  \LFIR     }{\ifmmode L_{\rm FIR} \else $L_{\rm FIR}$\fi}
\newcommand{  \Lfir     }{\ifmmode L_{\rm FIR} \else $L_{\rm FIR}$\fi}
\newcommand{  \LTIR     }{\ifmmode L_{\rm TIR} \else $L_{\rm TIR}$\fi}
\newcommand{  \Ltir     }{\ifmmode L_{\rm TIR} \else $L_{\rm TIR}$\fi}

\newcommand{  \mdyn     }{\ifmmode M_{\rm dyn} \else $M_{\rm dyn}$\fi} 
\newcommand{  \mgas     }{\ifmmode M_{\rm gas} \else $M_{\rm gas}$\fi} 
\newcommand{  \mh       }{\ifmmode M_{\rm h} \else $M_{\rm h}$\fi}
\newcommand{  \mhalo    }{\ifmmode M_{\rm halo} \else $M_{\rm halo}$\fi}
\newcommand{  \sfr      }{\ifmmode {\rm SFR} \else SFR\fi}

\newcommand{ \Lcii     }{\ifmmode L_{\cii} \else $L_{\cii}$\fi}
\newcommand{ \fwcii  }{\ifmmode {\rm FWHM}\cii \else FWHM\cii\fi}

\newcommand{\bj}{\ifmmode b_{\rm J} \else $b_{\rm J}$\fi}

\newcommand{\iab}{\ifmmode i_{\rm AB} \else $i_{\rm AB}$\fi}

\newcommand{\jab}{\ifmmode J_{\rm AB} \else $J_{\rm AB}$\fi}
\newcommand{\hab}{\ifmmode H_{\rm AB} \else $H_{\rm AB}$\fi}
\newcommand{\kab}{\ifmmode K_{\rm AB} \else $K_{\rm AB}$\fi}

\newcommand{\jveg}{\ifmmode J_{\rm Vega} \else $J_{\rm Vega}$\fi}
\newcommand{\hveg}{\ifmmode H_{\rm Vega} \else $H_{\rm Vega}$\fi}
\newcommand{\kveg}{\ifmmode K_{\rm Vega} \else $K_{\rm Vega}$\fi}

\newcommand{  \Chisq    }{\ifmmode \chi^{2} \else $\chi^{2}$}
\newcommand{  \nelec    }{\ifmmode n_{e} \else $n_{e}$\fi}     % electron density
\newcommand{  \nh       }{\ifmmode n_{\rm H} \else $n_{\rm H}$\fi}     % hydrogen density
\newcommand{  \Ncol     }{\ifmmode N_{\rm col} \else $N_{\rm col}$\fi} % column density
\newcommand{  \NH       }{\ifmmode N_{\rm H} \else $N_{\rm H}$\fi}     % column density

\def\sun{\hbox{$\odot$}}
\def\ion#1#2{#1$\;${\small\rm\@Roman{#2}}\relax}

\usepackage[latin1]{inputenc}
\usepackage[T1]{fontenc}

\received{November 10, 2021}
\revised{December 6, 2021}
\accepted{TBD}
\submitjournal{ApJ}

\shorttitle{AGNs and Reionization}
\shortauthors{Zeltyn \& Trakhtenbrot}
\graphicspath{{./}{figures/}}
\begin{document}

\title{The Contribution of AGN Accretion Disks to Hydrogen Reionization}

\author[0000-0002-7817-0099]{Grisha Zeltyn}
\affiliation{School of Physics and Astronomy, Tel Aviv University, Tel Aviv 69978, Israel}

\author[0000-0002-3683-7297]{Benny Trakhtenbrot}
\affiliation{School of Physics and Astronomy, Tel Aviv University, Tel Aviv 69978, Israel}

\correspondingauthor{Grisha Zeltyn}
\email{grisha.zeltyn@gmail.com,benny@astro.tau.ac.il}

\begin{abstract}

We examine the contribution of high-redshift ($z>6$) active galactic nuclei (AGNs) to cosmic hydrogen reionization, by tracing the growth and ionizing output of the first generation of supermassive black holes (SMBHs). 
Our calculations are anchored to the observed population of $z\simeq6$ quasars, and trace back the evolving spectral energy distributions (SEDs) of the accretion flows that power these early AGNs and consider a variety of growth histories, including super-Eddington accretion. 
Compared to a fixed-shape SED, the evolving thin disks can produce ionizing radiation that is higher by up to ${\sim}80\%$.
Across a variety of SMBH growth scenarios, the contribution of AGNs to reionization is limited to late epochs ($z<7$), and remains sub-dominant compared to star-forming galaxies.
This conclusion holds irrespective of the (still unknown) space density of low-luminosity $z=6$ AGNs, and for growth scenarios that allow super-Eddington accretion.
The contribution of AGNs to reionization can extend to earlier epochs ($z\gtrsim8$) in scenarios with relative slow SMBH mass growth, i.e., for low accretion rates and/or high spins.
We finally demonstrate that our framework can reproduce the observed quasar proximity zone sizes, and that compact proximity zones around $z=6$ quasars can be explained by the late onset of super-Eddington accretion.
\end{abstract}


%%%%%%%%%%%%%%%%%%%%%%%%%%%%%%%%%%%%%%%%%%%%%%%%%%%%%%%%%%%%%%%%

\section{Introduction}
\label{sec:intro}

The reionization of the Universe is the cosmic phase transition in which the hydrogen in the inter galactic medium (IGM), which was neutral ever since the recombination epoch ended at $z \sim 1100$, transformed into an ionized state. Recent {\it Planck}-based measurements of the IGM's Thomson optical depth to cosmic microwave background (CMB) photons suggest  $z_{\mathrm{0.5}}=7.68\pm0.79$ as the redshift at which half of the hydrogen was ionized, and points to a rather ``late'' reionization, occurring mostly at $z\lesssim10$ \citep{Planck18VI}. 
Complementary and independent measurements based on  Gunn-Peterson troughs in quasars \cite[e.g.,][]{Becker01,Djorgovski01,Fan03,Songaila04}, 
dark gap statistics in quasar spectra \cite[e.g.,][]{McGreer15}, and surveys of Ly$\alpha$-emitting galaxies \cite[e.g.,][]{Schenker14}, all point to reionization being (nearly) complete by $z\sim6$, with some recent studies suggesting that reionization has ended as late as $z\approx5.3$ \cite[e.g.,][]{Eilers18,Kulkarni19,Keating20,Bosman18,Bosman21,Zhu21}.

One key question concerning the reionization of the Universe is the nature of the sources of ionizing radiation that drive it. One type of such sources are young and hot stars in early star-forming galaxies. Multi-wavelength surveys of the high-redshift galaxy population 
have allowed to determine the evolution of the galaxy ultraviolet (UV) luminosity function at $z=6-10$ \cite[see, e.g.,][and references therein]{Stark16}, which, combined with assumptions about the ionizing emissivity and escape fraction of ionizing photons, can be used to calculate the production rate of ionizing photons by the high redshift galaxy population \cite[e.g.,][]{Robertson13,Bouwens15,Ishigaki18}. Such studies find that galaxies' ionizing radiation production rate is consistent with early galaxies being the main driver of reionization, with ionizing photon production rate densities at $z=6$ higher than $10^{50.5}\,\Nunits$ , which is higher than the production rate required to keep the IGM ionized at that redshift (assuming a clumping factor in the range of $2-4$; \citealt{Madau99}).

Another source of ionizing radiation are high redshift active galactic nuclei (AGNs) and quasars, powered by accreting super massive black holes (SMBHs). 
While quasars are far less common than star-forming galaxies, they are significantly more powerful sources of ionizing radiation.
The radiatively efficient accretion disks that power AGN produce high luminosities (reaching  $\Lbol\sim10^{47}\,\ergs$) over long timescales ($\gg10^6\,\rm{yr}$), with ionizing radiation emitted from their hot inner parts (which can reach temperatures of $\sim 10^5\,\rm{K}$). 
Over 480 quasars have been observed at $z>5$, 170 of which are at $z>6$ \cite[][and references therein]{RC20}, harboring SMBHs with masses at the range of $\sim10^{8-10}\,\Msun$ \cite[e.g.,][]{Willott10, Trakhtenbrot11, DeRosa14, Mazzucchelli17,Onoue19, Shen19}. The existence of these quasars, observed at the end of the epoch of reionization, makes them viable contributors to, or perhaps even the main drivers of, cosmic reionization.

One key outstanding issue related to the contribution of AGNs to reionization revolves around the uncertainty in the space density of low-luminosity ($\lesssim10^{45}\,\ergs$) AGNs at high redshifts, which are expected to be the most common among accreting SMBHs. 
The claim of a high number density of $z>4$ low-luminosity AGNs, by \citep[][followed by the more recent work \citealt{Giallongo19}]{Giallongo15}, combined with the {\it Planck}-based finding of a low Thompson scattering optical depth \citep{Planck18VI}, have resurfaced the idea that AGNs could have significantly contributed---or even dominated---cosmic hydrogen reionization \cite[e.g.,][]{MH15,Grazian20}. 
However, the \cite{Giallongo15} result was challenged by other studies \cite[e.g.,][]{Parsa18}, and it has become clear that it stands in some contrast with other observational efforts to identify $z\gtrsim5$, low-luminosity AGNs in appropriately deep surveys \cite[e.g.,][]{Weigel15,Cappelluti16,Akiyama18,Matsuoka18,McGreer18,Niida20}. 
Consequently, the space density of such systems and their contribution to reionization at $z\gtrsim6$ was claimed to remain limited \cite[e.g., ][]{Wang19,Shen20}.

The contribution of AGNs to reionization naturally depends on the output rate of ionizing photons from the accretion process, and on the corresponding spectral energy distribution (SED).
Many studies assume either a fixed-shape SED, or at least a narrow range of possible SEDs, motivated by observations in the relevant UV regime, \cite[e.g.,][]{Berk01,Telfer02,Lusso15}, and indeed across the EM spectrum \cite[e.g.,][]{Elvis1994,Marconi04}. 
While such SEDs can be scaled to fit different bolometric luminosities, they naturally result in the production rate of ionizing photons (denoted in this work as \Q) being a simple, often fixed fraction of the total photon production rate (i.e., the bolometric luminosity, \Lbol). 
In order to determine the total ionizing flux density of AGNs, these SEDs are combined with the quasar luminosity function (QLF), which at the relevant redshift regime often heavily relies on unobscured AGNs, i.e., quasars (i.e., the $z\sim6$ QLF; \citealt{Wang19,Shen20,Grazian20}). 
This pragmatic approach has two drawbacks.
First, as the QLF is currently unknown beyond $z\sim6$, it is extremely challenging to calculate the contribution of AGNs to reionization over the relevant, earlier period. 
In practice, this necessitates either extrapolating the (highly uncertain) redshift-dependent trends seen in the $z\lesssim6$ QLF \cite[e.g.,][]{HaimanLoeb98,Fontanot12, Grissom14, MH15, Garaldi19}, or relying on models of the (active) SMBH population \cite[e.g.,][]{BL01, TG11, Feng16, Qin17}.
Second, such assumptions do not take into account the possible variation in the SED, which is expected given the dependence of the accretion disk SED on the black hole (BH) parameters.
Specifically, in our work we focus on radiatively efficient, geometrically-thin, optically-thick accretion disks (\cite{SS73}), which show a good agreement to observed (low-$z$) AGN SEDs and spectra (see, e.g., \citealt{Davis07, Lusso15, Shang15, Capellupo15}, but see also \citealt{KB99, Jin12}). 
The high level of similarity between the broad-band SEDs \& UV spectra of $z\sim6$ quasars and those of lower redshift quasar samples \cite[e.g.,][]{Shen19,Vito2019_z6_Xrays,Pons2020_Xray_z65} suggests that the thin-disk model is also appropriate for the population of quasars at the end of the epoch of reionization.
As illustrated in Figure~\ref{fig:diff_SEDs}, for such thin-disk SEDs, \Q varies significantly with BH mass, accretion rate and spin\footnote{These dependencies are further discussed in Section~\ref{sec:methods_single_BH}.} -- all of which may evolve over the timescales relevant for the emergence of the first generation of SMBHs, and thus for their contribution to reionization. 
This variation and/or evolution cannot be captured by simple, essentially fixed-shape (power-law) SEDs.

Finally, a challenge that every reionization scenario focusing on AGNs has to address is the very formation and early growth of SMBHs. It is currently unknown how the observed $z\sim 5-7$ SMBHs have reached their high BH masses, over such a short period (i.e., $<10^9$ yr after the Big Bang; see, e.g., reviews by \citealt{Volonteri10, Volonteri12, VB12, Haiman13,LF16,JH16,Valiante17,Gallerani17,Inayoshi20}, and references therein).
One possible explanation involves stellar-mass BH seeds ($\lesssim10^2\,\Msun$) accreting at Eddington-limited accretion rates with a high duty-cycle. 
While possible in principle (as illustrated by, e.g., \citealt{Trakhtenbrot20}), accretion at such high duty cycles was shown to be unlikely in hydrodynamical simulations (see, e.g., \citealt{Inayoshi20} and references therein). Alternatively, several works have put forward the possibility of the formation of massive seeds ($10^4\lesssim M_{\rm seed}/\Msun \lesssim 10^6$; see, e.g., \citealt{KBD04, Begelman06, SS06} and reviews by \citealt{Natarajan2011,Volonteri12}). 
Another intriguing alternative explanation is that of stellar mass BH seeds accreting at super-Eddington rates, which can account for the observed $z=6$ SMBH masses even during short periods of accretion time \cite[e.g.,][]{Madau14,Volonteri15}. 
In the context of reionization it is important to stress that super-Eddington accretion could be highly luminous (i.e., consistent with the Eddington luminosity) even if radiatively inefficient \cite[e.g.,][]{McKinney14,SN16}.\footnote{This is in contrast to extremely low-rate accretion flows, which are both radiatively inefficient {\it and} extremely faint \cite[e.g.,][]{YuanNarayan14}.}
Note, however, that the observed accretion rates of $z\sim6$ quasars are consistent with high but sub-Eddington accretion \cite[e.g.,][]{Mazzucchelli17,Trakhtenbrot17,Onoue19,Shen19,Yang21}.

In our work we aim to investigate the contribution of AGNs  to reionization by interpreting the QLF at $z=6$ in a physical way, i.e., as a population of SMBHs with various SMBH-related parameters and corresponding, physically-motivated SEDs.
We then trace their growth back in time by considering various growth scenarios, while keeping track of their evolving SEDs and the corresponding ionizing radiation output. In Section~\ref{sec:methods} we present the methods used in this work to model the evolution of the ionizing output of a single SMBH, as well as the population of accreting SMBHs at high redshifts. 
Moreover, we incorporate a specific radiatively efficient  slim-disk model to account for super-Eddington accretion. Section~\ref{sec:results} presents the results obtained using these methods, followed by a detailed discussion, as well as a comparison of the ionizing output of SMBHs and galaxies, in Section~\ref{sec:dicussion}. We conclude with Section~\ref{sec:summary}, where we summarize the key results of our work and discuss its limitations and possible future extensions.

Throughout this work we assume a $\Lambda\mathrm{CDM}$ cosmology with $\Omega_m=0.3$, $\Omega_\Lambda=0.7$, $\Omega_b=0.047$, and $H_0=70\,\mathrm{km}\,\mathrm{s}^{-1}\mathrm{Mpc}^{-1}$.

\section{Methods}
\label{sec:methods}

\subsection{The ionizing output of a single SMBH}
\label{sec:methods_single_BH}

The main goal of this work is to evaluate the contribution of actively accreting $z\geq6$ SMBHs, powered by radiatively efficient accretion disks to cosmic reionization, by tracing an evolving population of AGNs, each contributing with an SED that is appropriate for the properties of the accreting SMBH.
To this end, we will mainly rely on two markedly different SED models:

\begin{itemize}
    \item A fixed-shape SED --- an observationally motivated SED based on the template constructed in \cite{Marconi04}. 
    This SED consists of a broken power-law ($L_\nu \propto \nu^{\alpha_\nu}$), with $\alpha_\nu = 2$ for $\lambda < 1\,\mic$ (the Rayleigh-Jeans tail of a blackbody),     $\alpha_\nu =-0.44$ in the range $1200\,\A < \lambda < 1\,\mic$, and $\alpha_\nu = -1.76$ in the range  $1\,\kev < \lambda < 1200\,\A$. 
    Since our aim is to model the accretion disk's SED we will ignore the part of the SED template blueward of 1 \kev, which is attributed to the hot corona. 
    This model has only one parameter --- $\Lbol$ --- which is used to normalize the SED, i.e. by requiring that $\Lbol\equiv\int L_\nu d\nu$.
    
    \item Standard thin disk --- the SED of a standard, \cite{SS73} like geometrically-thin, optically-thick accretion disk. 
    In particular, we construct these SEDs using a series of annular regions that follow the temperature profile as given in \cite{SS73}, and applying additional relativistic corrections \citep{NT73,PT74,RH95}. 
    In principle, this model has five parameters: The mass of the BH ($M$), the disk mass accretion rate (\Mdot), the dimensionless BH spin ($a$), and the inner and outer radii of the disk (\rin and \rout, respectively). 
    We take \rin to be the innermost stable orbit (ISCO), which is set by $a$. In addition, \rout is taken to be $500\,r_{\rm g}$ throughout this work. 
    Thus, in practice, the variety of SEDs we consider are determined by {\it three} parameters in total ($M$, \Mdot, and $a$). 
\end{itemize}

We assume a radiatively efficient accretion flow for both models. Thus, the bolometric luminosity, \Lbol, relates to the accretion rate through 

\begin{equation}
\label{eq:Lbol}
  \Lbol = \eta \Mdot c^2  \, ,
\end{equation}

where the radiative efficiency $\eta$ is a non-linear function of $a$.
In Section~\ref{sec:method-superEdd} below we describe an additional {\it slim}-disk model SED that is applicable in the super-Eddington regime.

For any given SED, the number of ionizing photons emitted per second, \Q, is given by
\begin{equation}
\label{eq:Q=integral}
    Q_{\rm ion} = \int\displaylimits_{13.6\,\mathrm{eV}}^\infty \frac{L_{\nu}}{h\nu}\mathrm{d}\nu\,,
\end{equation}
where the lower integration limit is set by the hydrogen ionization threshold frequency.

\begin{figure}
\centering
    \includegraphics[width=1.0\columnwidth]{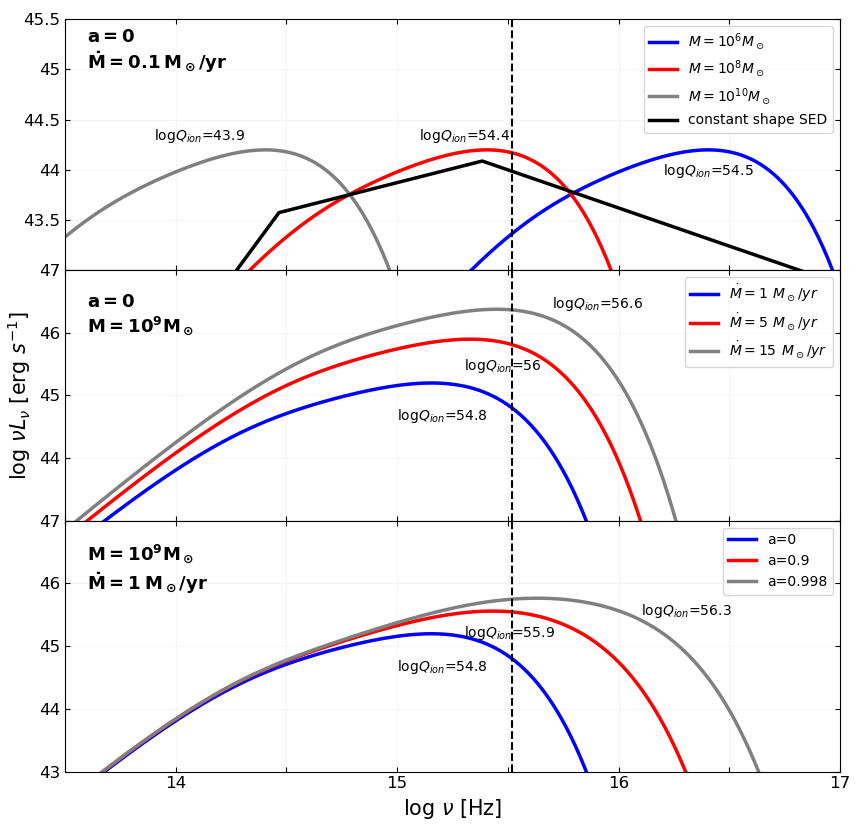}
    \caption{SEDs of standard geometrically-thin, optically-thick accretion disks with various parameters. Upper panel: all models have $a=0$ and $\Mdot=0.1\,\mpyr$, but various BH masses (the $M=10^{6}\,\Msun$ model slightly exceeds the Eddington limit). 
    Middle panel: various accretion rates, for $M=10^{9}\,\Msun$ and $a=0$. Lower panel: various BH spins, for $M=10^{9}\,\Msun$ and $\Mdot=1\,\mpyr$. 
    In all panels, the dashed vertical line marks the ionization frequency of hydrogen. For each SED, we note the corresponding value of \Q.
    The black solid line in the upper panel illustrates the fixed-shape SED used in this work (as well as other relevant studies), here normalized to the same \Lbol$\,$ as the rest of the SEDs in that panel.
    }
    \label{fig:diff_SEDs}
\end{figure}

Figure~\ref{fig:diff_SEDs} shows the SEDs of thin-disks for different BH parameters, with the appropriate \Q shown along each SED. 
The upper panel shows SEDs of non-spinning  BHs ($a=0$) with $\Mdot=0.1\,\mpyr$ and various BH masses. Note that changing the BH mass from $10^{10}$ to $10^{8}\,\Msun$ increases \Q by more than $10$ orders of magnitude, as a result of the accretion disk getting hotter for lower-mass black holes.
Decreasing the mass even further, to $10^6\,\Msun$, increases the disk temperature even still, but increases the number of emitted ionizing photons by only $\sim25\%$. 
This is due to the increase of the average energy of the emitted photons combined with the equal energy budget of each SED (Eq. \ref{eq:Lbol}). 
The middle panel in Figure~\ref{fig:diff_SEDs} presents SEDs for different accretion rates, where $M=10^9\,\Msun$ and $a=0$. In this case raising the accretion rate from $1$ to $15\,\mpyr$ increases \Q by a factor of ${\sim}60$, again due to the accretion disk having a higher temperature for higher values of accretion rates. In the lower panel the spin varies from a stationary BH to the maximal allowed spin for a thin accretion disk ($a=0.998$), for a $M=10^9\,\Msun$ BH with $\dot{M}=1\,\mpyr$, which results in an increase by a factor of ${\sim}30$ in \Q. 
For comparison, the upper panel shows the fixed-shape SED, normalized to have the same \Lbol\, as the other SEDs in this panel (i.e., $\Lbol\simeq3\times10^{44}\,\ergs$ for $\Mdot=0.1\,\mpyr$ and $a=0$). This clearly demonstrates that the constant shape SED, which can only be scaled, cannot capture the more complicated SED behavior as embodied by the standard thin-disk model, and specifically cannot fully account for the dependence of \Q on the BH parameters.

Note that the assumption of an isotropic radiation field (excluding the geometrical factor due to the disk's flat geometry) is generally not valid when relativistic effects are taken into account. These effects become more pronounced the closer the emission region is to the BH. Since for highly spinning BHs the ISCO, and hence the inner radius, are exceedingly small, the angle dependence of the emitted radiation will be stronger for high-spin BHs. To account for these relativistic effects we use the \texttt{KERRTRANS} code (presented in \citealt{Agol97}) to derive inclination angle dependent SEDs, and then integrate over all inclinations to derive the angle-integrated, frequency-resolved luminosity of the source (i.e., $L_\nu$). 
However, as demonstrated in Appendix~\ref{Appendix0} and Figure \ref{fig:kerrtrans vs regular}, for non-spinning BHs the differences between the SEDs calculated using \texttt{KERRTRANS} and those calculated while ignoring angle dependent relativistic effects are negligible. 
As the spins of high-redshift SMBHs are essentially unknown, or at least highly uncertain (e.g., \citealt{Trakhtenbrot17, Jones20}; see also the review by \citealt{Reynolds20} and references therein), we assume $a=0$ throughout most of this work. 
We therefore use SEDs generated using \texttt{KERRTRANS} only when exploring scenarios with highly spinning BHs.

In this work we consider several BH growth scenarios, which describe how the key BH properties evolve with (cosmic) time. These, in turn, determine the evolution of the SED, given an SED model (i.e., thin-disk or fixed-shape). In the growth scenarios considered in this work we take the BH spin to stay constant. Thus, the evolving SED would depend on the evolution of $M$ and \Mdot. A related quantity is the Eddington ratio, \fedd, defined as the ratio between the bolometric luminosity and the Eddington luminosity and which for radiatively-efficient (thin) accretion flows can be expressed as:
\begin{equation}
\label{eq:fedd}
    \fedd=\frac{\Lbol}{\Ledd}\propto\frac{\Mdot}{\dot{M}_{\rm Edd}}\propto\frac{\Mdot}{M}
\end{equation}

Currently, quasars are observed up to the early epochs when reionization is mostly complete ($z\simeq6-7.5$; \citealt{Banados18,Wang21}). 
Thus, when considering the evolution of a single SMBH, we will track its growth back in time, starting at the end-point of a $10^9\,\Msun$ SMBH at $z=6$ in order to be consistent with the typical observed mass at that redshift \cite[e.g,][]{TN12,DeRosa14,Trakhtenbrot17,Shen19}. 
Note, however, that we will explore various choices for \Mdot and $a$ at $z=6$, and the associated trends in BH growth and ioinizing output.
In our work we will rely on standard, thin-disk SEDs to consider two simple scenarios of BH growth:
\begin{itemize}
    \item 
    Constant $\fedd$ --- starting with the mass boundary condition at $z=6$, we evolve the BH parameters back in time at a fixed Eddington ratio. 
    This results in an exponential growth with a Salpeter timescale \citep{Salpeter64}, given by
    \begin{equation}
    \label{eq:salpeter}
        \tau_{\rm{Salpeter}}{\simeq}0.4\,\rm{Gyr}\,\frac{\eta}{(1{-}\eta)}\,\frac{1}{\fedd}.
    \end{equation}
    
    \item 
    Eddington-limited, constant $\Mdot\,$ (the \emph{``Mixed''} scenario hereafter) ---  starting with the mass boundary condition at $z=6$, we evolve the BH parameters back in time with a fixed physical accretion rate (i.e., in \mpyr). 
    Since growth at a constant \Mdot results in an increase in \fedd when going back in time (i.e. as $M$ decreases; Eq. \ref{eq:fedd}), the Eddington limit ($\fedd = 1$) will be reached at some time $t_\mathrm{Edd}$. Since the thin-disk models we consider are not valid in the super-Eddington regime, we will switch the growth for $t<t_\mathrm{Edd}$ to a growth at a constant $\fedd = 1$ (see above). 
    The physical motivation for such a scenario can be a galaxy that feeds mass at some average constant rate to the SMBH at its center, with the accretion rate itself being regulated by the Eddington limit.
\end{itemize}
We also explore a model that allows for accretion at super-Eddington rates (see Section~\ref{sec:method-superEdd} immediately below), thus enabling the BH to accrete at a constant \Mdot throughout its growth history (i.e., also at $t<t_\mathrm{Edd}$).

Since the SED depends on the disk's temperature range, which in turn depends on the mass and accretion rate, the time step in each of the above growth schemes is set so that the difference in mass between two consecutive steps is at most 10\%. 
This way the temperature difference between two consecutive steps does not exceed $\sim$5\%. For each time-step we calculate the SED and \Q of the appropriate  accretion disk, keeping track of their evolution as the BH grows.

\subsection{SEDs of super-Eddington slim disks}
\label{sec:method-superEdd}

Following the discussion in Section~\ref{sec:intro}, we are interested in modeling SMBH accreting at super-Eddington rates, for which the standard thin-disk model is inappropriate. While there are many different ideas for modeling super-Eddington accretion flows, in our work we focus on ``slim'',\footnote{The notion of ``slim'' disks commonly refers to a height-to-radius ratio of $H/R\sim1$, compared with $H/R \ll 1$ for ``thin'' disks (see \cite{Netzer13}).} luminous accretion disks, which have long been considered to be able to sustain super-Eddington flows onto BHs (at least at mild super-Eddington rates; e.g., \citealt{Abramowicz88}).

Specifically, we use the \texttt{AGNslim} model presented in detail in \cite{KD19}. We chose this specific model due to the availability of associated SED-generating tools. 
Detailing the physics behind the model, and/or its validity, are beyond the scope of this work, and the interested reader is referred to \citet[and references therein]{KD19} for a detailed discussion.
Here we briefly mention that \texttt{AGNslim} allows the disk to become slim due to increased radiation pressure, taking into account the increase in optical depth for high accretion rates, which causes `photon-trapping' near the mid-plane of the disk. These photons are then advected to the SMBH before they're able to escape (vertically) from the surface of the disk. In simple terms, this allows the disk to supply mass to the SMBH at super-Eddington rates while not violating the local Eddington limit near the disk surface. In addition, this model includes additional Comptonization which causes an excess of UV radiation. This model has $14$ parameters in total. 
Except for $M$, \fedd (which determines \Mdot) and $a$, we take all parameters to have the default values.\footnote{The full list of parameters can be viewed at \url{https://heasarc.gsfc.nasa.gov/xanadu/xspec/manual/node132.html}.} 
The \texttt{AGNslim} model does {\it not} consider outflows launched from the accretion flow \cite[e.g.,][]{DotanShaviv2011}.

\begin{figure}
    \centering
    \includegraphics[width=1.0\columnwidth]{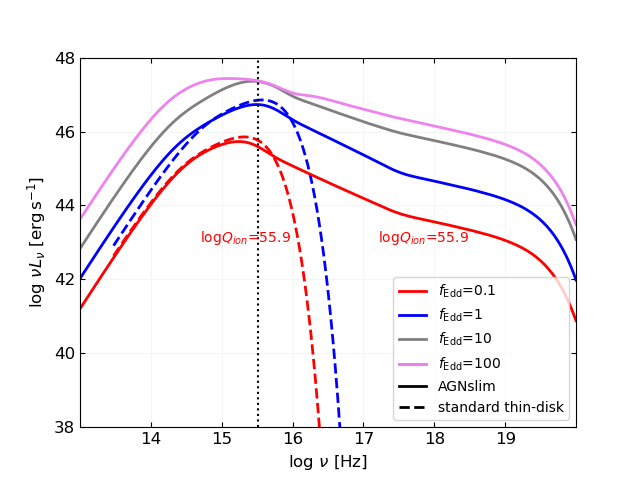}
    \caption{A comparison of the standard thin disk SED (dashed) and the slim disk SED produced by \texttt{AGNslim} (solid), for various choices of \fedd (colors). 
    The Lyman limit is marked by a vertical dotted line, and we note the values of \Q derived for the two SED types, assuming $\fedd=0.1$. 
    All the SEDs in this plot assume $M=10^9\,\Msun$ and $a=0$.
    The difference in \Q is negligible despite the apparent excess UV radiation in the \texttt{AGNslim} model, since many of the photons that are being further Comptonized in the slim disk ``envelope'' are already above the Lyman limit, and due to a slightly lower total number of photons produces by the slim disk.}
    \label{fig:SED_AGNslim_vs_thin}
\end{figure}

A comparison of the \texttt{AGNslim} SED and the standard thin disk SED, for various choices of \fedd, $M=10^9\,\Msun$ and $a=0$, is presented in Figure~\ref{fig:SED_AGNslim_vs_thin}. 
Note that although the slim disk SED is much harder than the standard disk SED, there is no significant difference in the number of ionizing photons produced by each SED. 
This is because, for these and similar parameters, many of the photons being Comptonized to higher energies in the \texttt{AGNslim} model are already above the Lyman limit, in addition to the slim disk producing a slightly lower total number of photons. The ionizing flux of the \texttt{AGNslim} model \textit{could} be significantly higher than that of the thin-disk case, for higher masses and/or lower accretion rates.

In order to calculate the ionizing photon production rate of a super-Eddington accreting flow we follow the same methods used for standard, Eddington-limited disks (described in the preceding Section~\ref{sec:methods_single_BH}). 
However, note that when the SED is described by the \texttt{AGNslim} model, the growth of the BH can proceed at a constant \Mdot and go over the Eddington limit at early times. Since the maximal allowed value for the Eddington ratio in this model is $1000\times\dot{M}_\mathrm{Edd}$, we will switch to a growth at a constant Eddington ratio equal to that value whenever this value is reached.
As we show in the next Section, the exact choice of this new maximal accretion rate does not affect our results, since the growth is already extremely fast.

\subsection{The ionizing output of a population of SMBHs}
\label{sec:methods_population}

In this work we use a bolometric QLF to describe the population of active SMBHs at $z=6$ (and beyond), which is parameterized as a double power-law of the form
\begin{equation}
\label{eq:double power-law}
\phi_{L}\left(L\right) = \frac{\phi^{\ast}}{\left(L/L^{\ast}\right)^{\gamma_{1}}+\left(L/L^{\ast}\right)^{\gamma_{2}}}\,,
\end{equation}
where $\phi^{\ast}$ is the comoving number density normalization, in units of $\rm{Mpc}^{-3}\,\rm{dex}^{-1}$; 
$L^{\ast}$ is the ``break'' luminosity; 
and $\gamma_{1}$ and $\gamma_{2}$ are the faint-end and bright-end power-law slopes, respectively. 
Specifically, for this work, we use QLFs from the recent study by \cite{Shen20}, as detailed in Section~\ref{sec:res_population}.

Our analysis aims to trace the active SMBH population, and thus the QLF, to $z>6$.
Instead of extrapolating the highly uncertain, empirical redshift-dependent trends seen at $z\sim4-6$ \cite[see detailed discussion in][]{Shen20}, we instead trace the evolution of the QLF based on a physically-motivated interpretation, coupled with our physical BH growth scenarios. 
Specifically, we associate every luminosity bin along the QLF with a certain set of BH parameters.  
We assume the same spin, and hence the same radiative efficiency, for the entire BH population. 
Thus, the bolometric luminosity at each point on the QLF will be associated with a certain accretion rate (Eq. \ref{eq:Lbol}).

In our work we take 30 BH mass bins, log-uniformly spread across the range $10^{6-10}\,M_{\odot}$ ($\Delta\log M\simeq0.13$ dex), to represent the reasonable mass range of the SMBH population at $z = 6$. 
In order to associate the QLF with this BH mass range we further assume a universal Eddington ratio for the entire active $z=6$ SMBH population. 
This assumption allows us to effectively deduce an active SMBH mass function (BHMF) from the QLF, with the most massive BH being the most luminous and rarest systems. 
We can then use the mass growth schemes described in Sections~\ref{sec:methods_single_BH} and ~\ref{sec:method-superEdd} to trace back the evolving QLF and BHMF for any $z>6$.\footnote{Note that the SMBH masses at $z>6$ will be necessarily smaller than those imposed at $z=6$, i.e. there will be no SMBHs with $M=10^{10}\,\Msun$ at $z>6$.}

The space density of quasars as a function of \Q can be derived from the QLF by
\begin{equation}
\label{eq:L-to-Q density}
\phi_{Q}\left(Q_{\rm ion}\right) = \phi_{L}\left(L\right)\frac{\Delta \log L}{\Delta \log Q_{\rm ion}}\,,
\end{equation}
where $\Delta\log L$ and $\Delta\log Q_{\rm ion}$ are small logarithmic intervals of bolometric luminosity and  production rate of ionizing photons, respectively, associated with consecutive BH mass bins within the (evolving) mass range.

By integrating the $Q_{\rm ion}$ distribution over the entire range of masses, at any specific redshift, we get the ionizing photon flux density of the redshift-resolved progenitor population of the $z=6$ SMBHs:
\begin{equation}
\label{eq:Nion}
\dot{N}_{\rm ion} = \int^{Q_{\rm ion}\left(M_{\rm max}\right)}_{Q_{\rm ion}\left(M_{\rm min}\right)} \phi_{Q}\left(Q_{\rm ion}\right) \mathrm{d}Q_{\rm ion}\,.
\end{equation}
This quantity can be directly compared to the ionizing photon flux needed for reionizing the Universe and/or to the output of other ionizing sources (i.e., early galaxies).

%%%%%%%%%%%%%%%%%%%%%%%%%%%%%%%%%%%%%%%%%%%%%%%%%%%%%%%%%%%%%%%%
\section{Results}
\label{sec:results}

In this Section we rely on the methods outlined above to trace the ionizing output of a single SMBH through $z=6-20$ under several Eddington-limited growth scenarios, and assess the output of the evolving SMBH population. 
We then re-examine the radiative outputs when super-Eddington growth is allowed.

\subsection{How many ionizing photons does a single SMBH produce?}
\label{sec:res_single_bh}

In this section we will present the evolution of \Q for growing BHs with various parameters and growth schemes, as detailed in Section~\ref{sec:methods_single_BH}. In all the scenarios outlined in this section the BHs reach a mass of $\mbh=10^9\,\Msun$ at $z=6$.

\begin{figure}[t]
    \centering
    \includegraphics[width=1.0\columnwidth]{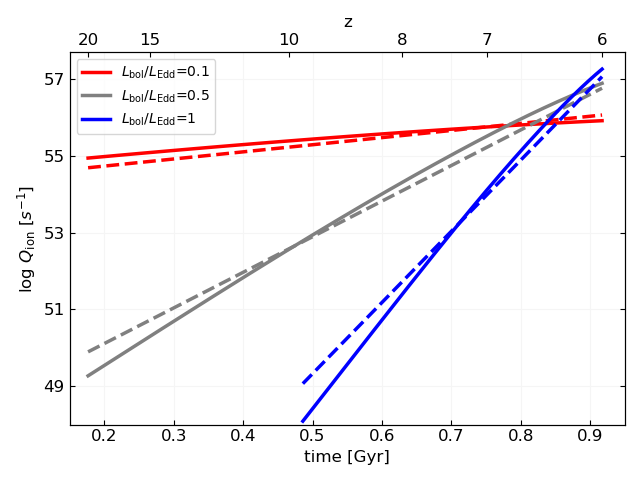}
    \caption{Evolution of \Q for a spinless BH growing through a constant Eddington ratio accretion, for different choices of \fedd. 
    Calculations with a standard thin-disk model are presented in solid lines and those with a fixed-shape SED in dashed lines. 
    By construction, all growth scenarios result in a $10^9\,\Msun$ SMBH at $z=6$. The differences in \Q at this ``final'' epoch are due to the different SEDs associated with each of the models.
    } % caption
    \label{fig:Qvst_const_fEdd}
\end{figure}

Figure~\ref{fig:Qvst_const_fEdd} presents the evolution of \Q over $6 \leq z \leq 20$ for spinless BHs with three different choices of a {\it constant} Eddington-ratio,  $\fedd=0.1, 0.5$ and $1$. 
We show calculations carried out both with the standard, thin-disk  model (solid lines) and with the fixed-shape SED model (dashed lines). 
Note that the various models do {\it not} result in identical ionizing outputs at $z=6$, as the different choices of \fedd\ and SED model would necessarily change \Q.
In the latter fixed-shape SED scenario, the constant \fedd\ leads to an exponential growth in \Lbol\ (Eq. \ref{eq:salpeter}), and thus we expect an exponential growth in \Q. 
We expect the standard thin-disk to result in a more complex behavior, since now not only \Lbol\ is changing but also the shape of the SED. As Figure~\ref{fig:Qvst_const_fEdd} shows, however, this model still results in an approximately exponential growth. 
By integrating the number of photons over time, we get the total number of ionizing photons produced by the BH during its entire growth. For our choice of parameters, this total time-integrated number of ionizing photons for the thin-disk model is higher by a factor of ${\sim}1.6$ and ${\sim}1.75$ compared to the fixed-shape SED model for $\fedd=0.5$ and $1$, respectively. For $\fedd=0.1$ there is no significant difference in the total ionizing output between the two models.

\begin{figure*}
    \centering
    % \begin{subfigure}
    % \centering
    \includegraphics[width=0.48\textwidth]{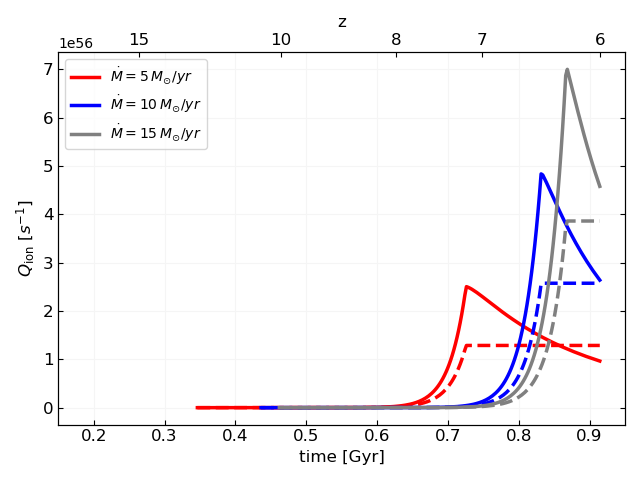}
    \hfill
    \includegraphics[width=0.48\textwidth]{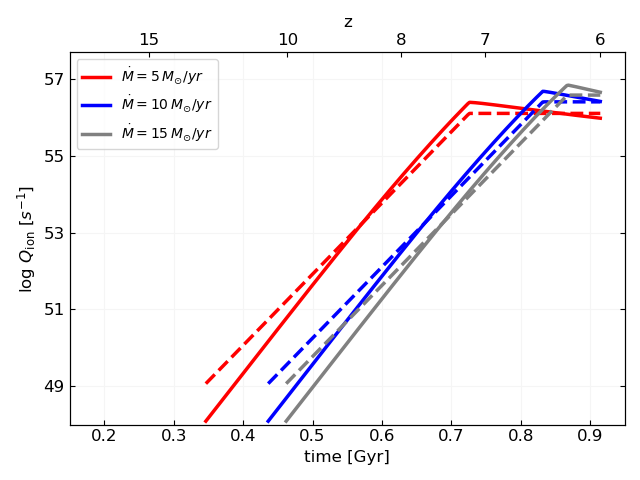}
    % \end{subfigure}
    \caption{\textit{Left}: Evolution of \Q in linear-scale for a spinless BH growing through a constant, Eddington-limited \Mdot accretion, for various choices of \Mdot. 
    Calculations with a standard thin-disk model are presented in solid lines and those with a fixed-shape SED in dashed lines. 
    All growth scenarios result in a $10^9\,\Msun$ SMBH at $z=6$. \textit{Right}: Same but \Q is in log-scale.
    } % caption
\label{fig:Qvst_Mixed}
\end{figure*}

%%% 
For Eddington limited accretion with a constant \Mdot the difference between the two models is more striking. 
The evolution of \Q for this ``Mixed'' growth scheme, for spinless BHs with different choices of $\Mdot(z=6)$, is presented in Figure~\ref{fig:Qvst_Mixed}, where we again show both thin-disk and fixed-shape calculations, with \Q plotted in linear scale on the left and log-scale on the right. 
In this growth scenario both SED models exhibit two distinct regimes of \Q evolution: 
\begin{enumerate}
    \item 
    At late times, starting at $z=6$ and going backwards in time, accretion proceeds with a constant, sub-Eddington \Mdot, resulting in a constant \Q for the fixed-shape SED and a \Q that decreases with increasing time for the thin-disk SED. 
    This latter trend is a unique consequence of the thin-disk SED model, and is driven by the increase in disk temperature, and thus in ionizing photons, with decreasing BH mass. Note that this trend is {\it not} caused by an increase in the luminosity, but rather by the changes to the SED (see  Fig.~\ref{fig:SED_evolution_demo}).
    \item 
    At earlier times, when the system reaches its Eddington limit, the BH can no longer maintain the assumed $\Mdot(z=6)$ and thus the accretion switches to a constant $\fedd=1$. In this case the change in \Q is, again, exponential for both SED models.
    \item 
    The transition between the late, constant \Mdot regime and the early, constant \fedd\ regime for the thin-disk model manifests itself in a pronounced peak in \Q. This peak occurs at the time and BH mass in which the assumed $\Mdot\left(z=6\right)$ corresponds to the Eddington limit (with the latter quantity being mass- and thus time-dependent). 
\end{enumerate}
The Eddington-limited, constant \Mdot scenario, coupled with the thin-disk spectral evolution, produces a factor of ${\sim}2$ more ionizing photons at the peak of emission, as compared to the fixed-shape SED model. 
The total, time-integrated, number of ionizing photons produced by the former model is higher by a factor of ${\sim}1.6$, ${\sim}1.5$ and ${\sim}1.3$ compared to the latter, for \Mdot of $5$, $10$ and $15$ $\Msun/\rm{yr}$ at $z=6$, respectively.
Figure~\ref{fig:SED_evolution_demo} (in Appendix \ref{Appendix1}) demonstrates the time-evolving SEDs calculated for one of the growth scenarios depicted in Figure~\ref{fig:Qvst_Mixed} (the blue curve, with $\Mdot=10\,\mpyr$ at $z=6$). 
At late times, the SEDs evolve ``horizontally'' due to the increasing mass, causing a decrease in \Q, while in earlier time, during the constant Eddington ratio accretion growth phase, the SEDs evolve both ``horizontally'' and ``vertically'', causing an exponential increase in \Q.

\begin{figure}
\centering
    \includegraphics[width=1.0\columnwidth]{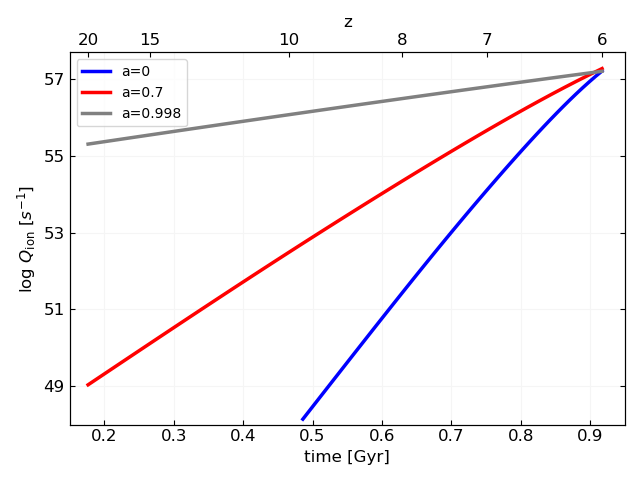}
    \caption{Evolution of \Q for a BH accreting at a constant $\fedd=1$ for various choices of (constant) BH spin ($a$). 
    All growth scenarios result in a $10^9\,\Msun$ SMBH at $z=6$.}
\label{fig:Qvst_diff_spin}
\end{figure}

Another parameter that affects \Q is the BH spin. 
Figure~\ref{fig:Qvst_diff_spin} presents the evolution of \Q for different choices of constant BH spin, where the mass growth proceeds at a constant $\fedd=1$. 
Note that a more realistic model should take into account the spin evolution during steady and/or intermittent accretion episodes \cite[e.g.,][see discussion in Section~\ref{sec:discussion_extending}]{King08,Dotti13}. 
Nonetheless, our simplistic assumption of a constant spin can still demonstrate the basic effects that different spin values have on the accretion disk's production rate of ionizing photons. We recall that higher BH spins provide higher radiative efficiencies. 
This results in a total, time-integrated, higher number of ionizing photons for high spinning BHs, 
with the $a=0.998$ BH producing a factor of ${\sim}6$ higher \Q than a spinless BH of an identical mass and accretion rate. 

One important caveat to this result is that higher spins imply higher radiative efficiencies, and thus slower mass growth.\footnote{The assumption of a fixed \fedd also implies that a higher spin (i.e., higher $\eta$ in Eq.~\ref{eq:Lbol}) BH will have a lower \Mdot, which is another contributor to its slower mass growth.} 
Assuming a certain final mass at $z=6$, this causes the \textit{implied} masses we get at any given earlier epoch to be higher for highly spinning BHs, also implying higher BH seed masses (see discussion in Section~\ref{sec:discussion_extending}).

\subsection{The evolution of the ionizing flux density of SMBHs}
\label{sec:res_population}

In this section we will follow the methods described in Section~\ref{sec:methods_population} to trace the evolution of the QLF, and the corresponding \Q density distribution function, from $z=6$ to earlier epochs. We will then derive the ionizing flux density, \Nion, of the entire population of actively accreting SMBHs over the range $z=6-15$.

Since there is currently a lack of observations of faint AGNs at $z=6$, we adopt two different QLFs which differ mainly in their faint-end slopes in order to represent the uncertainty in this regime. In the present work we use QLFs that were constrained by multiwavelength observations and presented in the recent paper by \cite{Shen20}. 
Specifically, we use the ``global fit A'' and ``B'' QLFs in \citet[see their Fig.~5]{Shen20}---hereafter referred to simply as ``QLF A'' and ``QLF B'', respectively.
``QLF A'' corresponds to a considerably higher number of low-luminosity AGNs at $z\sim6$, compared to ``QLF B''.
At the faint end, ``QLF A'' is comparable to the QLF derived by \cite{Giallongo19}. 
For example, at $\Lbol=10^{44}\,\ergs$, both the QLF denoted as ``Model 3'' in \cite{Giallongo19} and ``QLF A'' give $\phi_L\simeq4.8\times10^{-5}\,{\rm dex}^{-1}\,{\rm cMpc}^{-3}$, which is a factor $\sim20$ higher than the corresponding value for ``QLF B''.
We thus use ``QLF A'' to represent the most ``optimistic'' case in terms of the number of faint AGNs at $z>6$. 
The parameters of the QLFs we consider are listed in Table \ref{table:QLF parameters}.

\begin{figure*}
    \centering
    % \begin{subfigure}
    % \centering
    \includegraphics[width=0.48\textwidth]{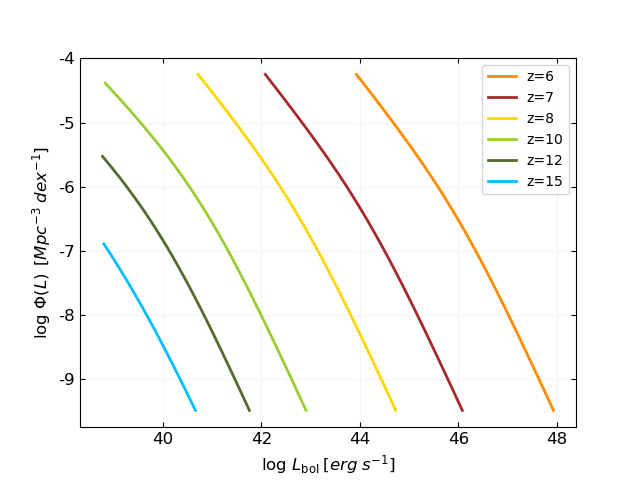}
    % \end{subfigure}
    % \begin{subfigure}
    % \centering
    \hfill
    \includegraphics[width=0.48\textwidth]{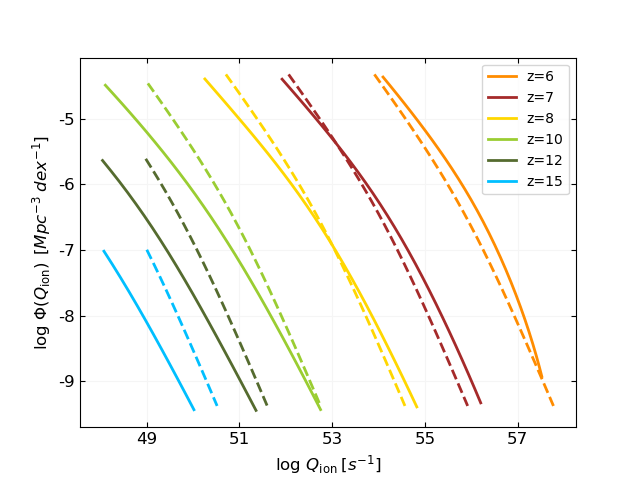}
    % \end{subfigure}
    \caption{The evolution of the QLF (left) and of the \Q density function (right) between $z=6$ to $z=15$.
    Here we assume constant $\fedd=0.6$ and $a=0$ for the entire population of SMBHs, and ``QLF A'' at $z=6$. Solid lines mark calculations with the standard thin disk SED and dashed lines represent the fixed-shape SED model.}
\label{fig:QLF_ evo}
\end{figure*}

An example of the evolution of a QLF from $z=6$ to $z=15$ is shown in the left panel of Figure~\ref{fig:QLF_ evo}. 
Here we used ``QLF A'' at $z=6$, and the Eddington ratio and spin of the entire SMBH population were fixed at $\fedd=0.6$ and $a=0$, respectively.
% , and both were held constant with time. 
In this case, the standard thin disk and the constant shape SED produce almost identical curves, so only the standard thin disk curves are shown. Figure~\ref{fig:QLF_ evo} shows that under the above assumptions, the QLF exhibits a pure luminosity evolution, meaning that the evolving QLF can be described solely by the evolution of $L^*$. 
In contrast, as can be seen in the right panel of Figure \ref{fig:QLF_ evo}, the resulting evolution of the \Q density function differs for the two SED models, with the thin-disk model \Q density function becoming ``steeper'' the lower the redshift, as well as displaying a faster-evolving ``horizontal shift'', than with the fixed-SED model.

\begin{figure}
\centering
    \includegraphics[width=1.0\columnwidth]{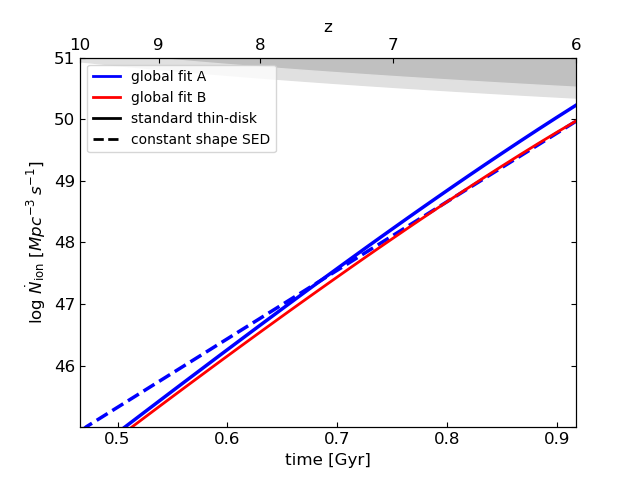}
    \caption{The evolution of \Nion derived for the population of accreting SMBHs, for different QLFs.
    All calculations assume that all SMBHs accrete with constant $\fedd=0.6$ and $a=0$. 
    Blue and red lines represent $z=6$ ``QLF A'' and ``B'' from \cite[][respectively]{Shen20}. In each set of calculations, solid lines represent the thin disk model while dashed lines represent the fixed-shape SED model. 
    Gray-shaded regions are the ionizing flux density required to keep the Universe ionized, with the lower and upper boundaries of the light-gray region corresponding to a clumping factor of $C=2$ and $C=3$, respectively.}
\label{fig:Qvsz}
\end{figure}

The ionizing flux density of the entire SMBH population at any given redshift can be obtained by integrating the \Q density function at that redshift (Eq. \ref{eq:Nion}) over the range of \Q. Figure \ref{fig:Qvsz} shows the result of this calculation, assuming $\fedd=0.6$ and $a=0$ as before. Here we plot \Nion for the standard thin-disk (solid blue) and the fixed-shape  (dashed blue) SED models, assuming ``QLF A'' at $z=6$. 
The resulting \Nion at $z=6$ for the thin-disk model is higher by a factor of $\sim1.8$ than that of the fixed-shape SED model. Moreover, the total number of ionizing photons emitted (per $\mathrm{Mpc}^3$) over the entire redshift range for the thin-disk model is higher than the fixed-shape one by a factor of $\sim1.75$. These results are consistent with the results obtained for a single SMBH (Section~\ref{sec:res_single_bh}), which showed that accounting for the spectral evolution of the SED increases the ionizing output by similar factors.

Figure~\ref{fig:Qvsz} also shows, in solid red line, a similar calculation done with the thin-disk SED model and ``QLF B'' (instead of ``A''). 
As this QLF has far fewer  low-luminosity AGNs, the region between the two solid lines in Figure~\ref{fig:Qvsz} reflects the underlying uncertainty in the ionizing flux density produced by the AGN population due to the unknown number of such low-luminosity AGNs at $z\gtrsim6$. In this regard, we recall that both QLF models from \cite{Shen20} used in this work are essentially unconstrained by actual measurements of low-luminosity ($L\lesssim10^{45}\,\ergs$) AGNs at $z\sim6$ (see their Fig.~5). 
At $z=6$, this uncertainty gives rise to a factor of $\sim 1.8$ higher ionizing flux density for the higher number of faint AGNs scenario. Integrated in time, the population with the more numerous number of faint AGNs produces a factor of $\sim 1.7$ higher total number of ionizing photons (per $\mathrm{Mpc}^3$) than the population with the lower number of faint AGNs. This minor difference in the ionizing flux density between the two populations, as compared to the seemingly significant difference in faint AGNs (a factor of $\sim20$ for AGNs powered by a SMBH with $M
\simeq 10^6\,\Msun$), is discussed in more detail in Section~\ref{sec:faint AGNs}.

\begin{deluxetable}{c|cccc}
\tablenum{1}
\label{table:QLF parameters}
\tablecaption{The QLF parameters used in this work}
\tablewidth{0pt}
\tablehead{
\colhead{Name} & \colhead{$\log\,\phi^{\ast}$} & \colhead{$\log\,L^{\ast}$} &
\colhead{$\gamma_1$} & \colhead{$\gamma_1$} \\
\colhead{} & \colhead{$[\mathrm{dex}^{-1}\,\mathrm{cMpc}^{-3}]$} & \colhead{$[\mathrm{erg}\,\mathrm{s}^{-1}]$} &
\colhead{} & \colhead{}
}
\startdata
QLF A & 2.702 & 46.06 & 0.9671 & 1.694 \\
QLF B & 2.962 & 45.91 & 0.2196 & 1.699 \\
\enddata
\tablecomments{See \cite{Shen20} for more details.}
\end{deluxetable}

To put our calculations in the context of cosmic (hydrogen) reionization, Figure~\ref{fig:Qvsz} finally shows the ionizing flux required to keep the Universe ionized as a function of redshift (as a gray-shaded region). 
The lower (upper) boundary of the region shown corresponds to a clumping factor of $C=2$  \cite[3, respectively; see][]{Madau99}.

\begin{figure}
\centering
    \includegraphics[width=1.0\columnwidth]{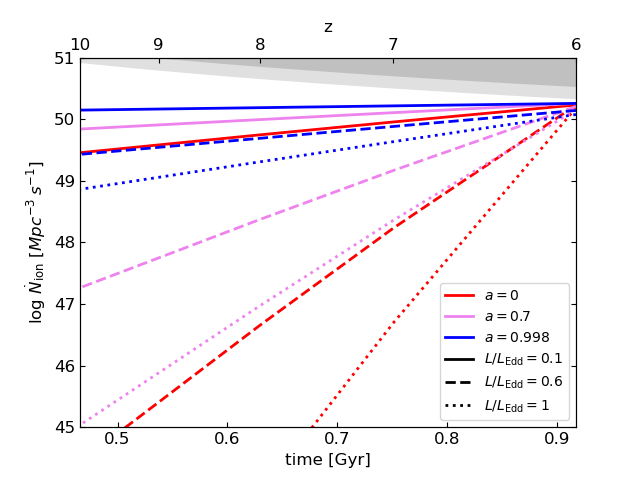}
    \caption{The evolution of \Nion derived for the population of accreting SMBHs, for different values of spin and \fedd, assuming accretion at a constant Eddington ratio. All calculations assume $z=6$ ``QLF A''. 
    Gray-shaded regions are the ionizing flux density required to keep the Universe ionized (as in Fig.~\ref{fig:Qvsz}).}
\label{fig:Nion_diff_spin_fEdd}
\end{figure}

Figure \ref{fig:Qvsz} shows that for our particular choices of fixed $\fedd=0.6$ and $a=0$ the contribution of SMBHs to reionization can be significant only at late stages, near $z\simeq6$, with \Nion dropping sharply by $2$ orders of magnitude already by $z=7$. The impact of different choices of parameters on the contribution of SMBHs to reionization is explored further in Figure \ref{fig:Nion_diff_spin_fEdd}, which shows the evolution of \Nion for different BH spins and Eddington ratios, and assuming the fiducial ``QLF A''. These parameters determine the steepness of the curve, with higher spins and lower (universal) Eddington ratios showing ``flatter'' curves and thus a possible extension of the contribution of accreting SMBHs to reionization to earlier times (higher $z$). 
These trends are mainly driven by the fact that fast-spinning BHs have higher radiative efficiency, causing the BH to grow more slowly, maintaining a high ionizing output during their growth up to $z=6$. 
Similar considerations explain the effect of changing \fedd: a lower \fedd means a lower \Mdot to $M$ ratio, which results in a slower mass growth.

\subsection{How does super-Eddington accretion affect the ionizing output?}
\label{sec:res_supEdd}

In this section we apply the methods outlined in Section~\ref{sec:methods_single_BH} and Section~\ref{sec:methods_population} to the \texttt{AGNslim} SED model, described in Section~\ref{sec:method-superEdd}, to account for the possibility of super-Eddington accretion and fast growth of the first generation of SMBHs. 
Throughout this work we will refer to this model simply as the ``slim disk'' or the ``super-Eddington'' model. 
In what follows, we will use  \MMedd instead of \fedd (i.e., \MMedd\ instead of \lledd), since in super-Eddington accretion flows \Lbol\ saturates and thus no longer scales linearly with \Mdot (as in Eq. \ref{eq:Lbol}; see discussion in Section~\ref{sec:discussion_superEdd}).

\begin{figure}
\centering
    \includegraphics[width=1.0\columnwidth]{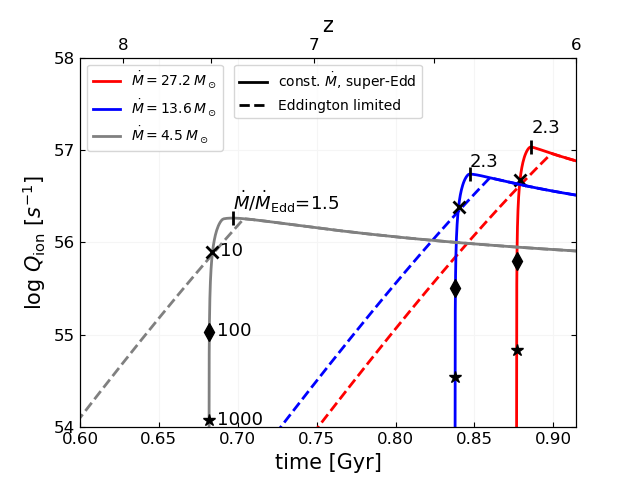}
    \caption{The evolution of \Q for the slim disk model, using \texttt{AGNslim}. 
    Solid lines represent super-Eddington growth scenarios, where accretion proceeds at a constant \Mdot that can exceed the Eddington limit. 
    Dashed lines represent the corresponding Eddington-limited growth scenarios, where the accretion follows the same (constant) \Mdot but only as long as it does {\it not} exceed the Eddington limit. 
    The different colors represent different choices of $\Mdot\left(z=6\right)$.
    For the solid curves, \MMedd at the peak of ionizing output emission is marked by a black vertical line, and the points where $\dot{M}/\dot{M}_{\rm Edd}$ reaches the values of $10$, $100$ and $1000$ are marked by different symbols. All the calculations shown here assume $M\left(z=6\right)=10^9\,\Msun$ and a constant $a=0$.}
\label{fig:Qvst_AGNslim}
\end{figure}

Because the slim disk model is valid in the super-Eddington regime, it allows to describe BHs that accrete at a constant \Mdot throughout their entire mass growth. Figure \ref{fig:Qvst_AGNslim} shows the ionizing output histories of BHs with $a=0$ and $M\left(z=6\right)=10^9\,\Msun$ for different values of $\Mdot\left(z=6\right)$. 
We show growth scenarios with accretion fixed at a constant \Mdot that may exceed the Eddington limit (in solid lines), as well as the corresponding Eddington-limited scenarios (dashed lines), which we have previously described as the ``Mixed'' growth scenario. 
Starting at $z=6$ and going backwards in time, both accretion modes produce similar, increasing ionizing outputs as long as the accretion is below the Eddington limit. When the Eddington limit is reached, the two sets of curves diverge: the Eddington-limited accretion stays at $\MMedd=1$, resulting in an exponential evolution of mass and \Q, as discussed in Section~\ref{sec:res_single_bh}, while the slim-disk constant \Mdot accretion continues with further increase in \MMedd and in \Q. This increase is not maintained for long, with \Q peaking at $\MMedd \approx1.5-2.3$. 
Beyond this peak (i.e., earlier times), the ionizing output drops sharply---by more than two orders of magnitude within $\Delta t\sim10\,\mathrm{Myr}$---while \MMedd continues to increase, until the BH reaches the seed mass of $10\,\Msun$. 
This sharp change in \Q is caused by the combined effect of (1) advection of ionizing photons in the super-Eddington regime, and (2) the more generic decrease in ionizing output with decreasing BH mass, which is indeed very dramatic in the super-Eddington regime (see further discussion in Section~\ref{sec:discussion_superEdd}).

Note also, that despite the apparently significant differences between the Eddington-limited, ``Mixed'' scenario and the super-Eddington, constant \Mdot scenario seen in Figure~\ref{fig:Qvst_AGNslim}, the total (time integrated) number of ionizing photons produced in the former scenario is higher by only $\sim$10\% than in the latter.

\begin{figure}
    \centering
    \includegraphics[width=1.0\columnwidth]{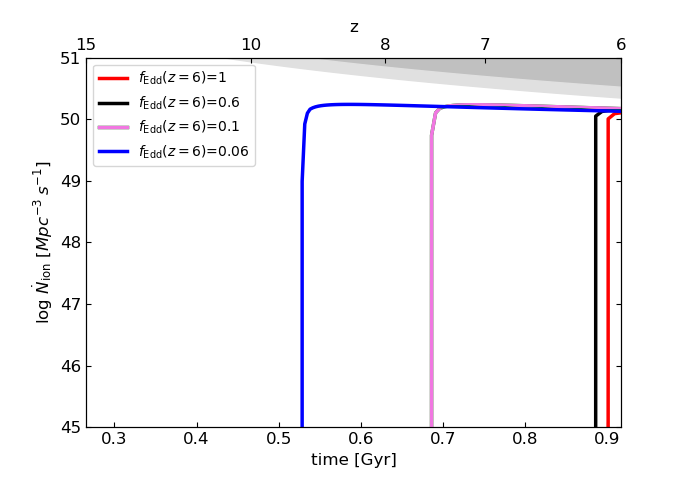}
    \caption{The evolution of the total ionizing flux density of all accreting SMBHs, whose SEDs are described by the slim disk model, where the accretion proceeds with a constant \Mdot. \Mdot for each mass interval is set by the value of \fedd at $z=6$ (represented in various colors). 
    We also assume $a=0$ and ``QLF A'' at $z=6$. 
    Gray-shaded regions are the ionizing flux required to keep the Universe ionized (as in Fig.~\ref{fig:Qvsz}).}
    \label{fig:Nion AGNslim}
\end{figure}

The slim disk model can be also applied to the population analysis as outlined in Section~\ref{sec:methods_population}. Figure~\ref{fig:Nion AGNslim} shows the evolution of \Nion for a population of (spinless) SMBHs, which  is described by ``QLF A'' at $z=6$ from \cite[see Table \ref{table:QLF parameters}]{Shen20} and by different values of $\fedd\left(z=6\right)$, setting the accretion rate for each BH mass interval. 
Going backwards in time, the accretion of all BHs proceeds at a constant \Mdot. This results in the \Nion curves looking almost like a step function: the SMBH population produces a steady, relatively high supply of ionizing photons ($>10^{50}\,\Nunits$) at late times and is sharply terminated at a certain redshift. This ``termination redshift'' is higher for lower values of $\fedd(z=6)$, for similar reasons to the ones discussed in Section~\ref{sec:res_population}: a lower \fedd (i.e., a lower $\Mdot/M$ ratio) translates to slower mass growth (while maintaining the same total integrated space luminosity density at $z=6$). 
This result emphasizes that the contribution of accreting SMBHs to the reionization of the Universe can be significant at late times ($z\sim6$), and can be extended to higher redshifts by means of relatively low (typical) accretion rates.

%%%%%%%%%%%%%%%%%%%%%%%%%%%%%%%%%%%%%%%%%%%%%%%%%%%%%%%%%%%%%%%%
\section{Discussion}
\label{sec:dicussion}

In what follows we discuss in more detail several aspects of our results. First, in Section~\ref{sec:faint AGNs}, we discuss the impact of the unknown number of faint AGNs at $z=6$ on \Nion. Next, in Section~\ref{sec:discussion_superEdd}, we discuss some insights related to the ionizing output of super-Eddington, slim accretion disks. We then discuss how various growth scenarios can extend the contribution of SMBHs to reionization to higher redshifts in Section~\ref{sec:discussion_extending}. We follow with a comparison between the ionizing flux density of SMBHs and galaxies in Section~\ref{sec:discussion_galaxies}. Finally, in Section~\ref{sec:discussion_ionized_regions}, we explore the implications of evolving SEDs on the sizes of proximity zones around quasars, by performing simplistic (1D) IGM radiative transfer calculations.

\subsection{The contribution of faint AGNs to reionization}
\label{sec:faint AGNs}

In Figure~\ref{fig:Qvsz} we show calculations of the ionizing flux density of AGNs using two $z=6$ QLFs: ``QLF A'' (blue) and ``B'' (red). These two QLFs differ mainly in the number of faint AGNs, with the former having an AGN density at $\Lbol =10^{44}\,\ergs$ that is higher by a factor of $\sim20$ than that of the latter.
In light of this, our result that the difference in \Nion at $z=6$ between the two QLFs is only ${
\sim}80\%$ (see Section~\ref{sec:res_population}) may seem surprising. 
This contrast can be explained by considering that (1) the {\it total} space density of AGNs, across $10^{44} < \Lbol/\ergs < 10^{48}$, for the QLF with the high number of faint AGNs is higher only by a factor of ${\sim}7$ compared to the other QLF; 
and (2) that the additional AGNs in the higher-density QLF indeed have relatively low luminosities ($\Lbol \lesssim 10^{45}\,\ergs$) and  thus relatively low \Q, as compared to the higher luminosity AGNs. 
In other words, the bulk of the ionizing radiation for both QLFs is produced by the AGNs in the high luminosity regime, where the two QLFs are rather similar, being well constrained by observations.
Lower luminosity AGNs, irrespective of their abundance, do not contribute significantly to the \Nion of the whole AGN population.
We postulate that the difference in \Nion will be more significant for QLFs with a higher number of {\it intermediate} luminosity AGNs (i.e., if $\phi_\star$ and/or $L_\star$ are higher; see Table \ref{table:QLF parameters}). 
\citealt{Giallongo15, Giallongo19} has indeed derived such a QLF based on observations that are focused on $4.5 \lesssim z \lesssim 5$, but there is currently no robust evidence that the $z=6$ QLF has comparable features \cite[e.g.,][]{Shen20, Kim20, Niida20}.

\subsection{Super-Eddington accretion}
\label{sec:discussion_superEdd}

We next turn to the growth scenarios that involve super-Eddington accretion. Figure~\ref{fig:Qvst_AGNslim} shows the evolution of ionizing photon flux for growth scenarios that allow for super-Eddington accretion through slim disks.
It demonstrates that when looking back in time, super-Eddington accretion produces a very sharp drop in \Q, of over 2 dex, within less than 10 Myr, when the SMBH reaches (mass accretion) Eddington ratios of ${\sim}1.5{-}2.3$. 

\begin{figure}
\centering
    \includegraphics[width=1.0\columnwidth]{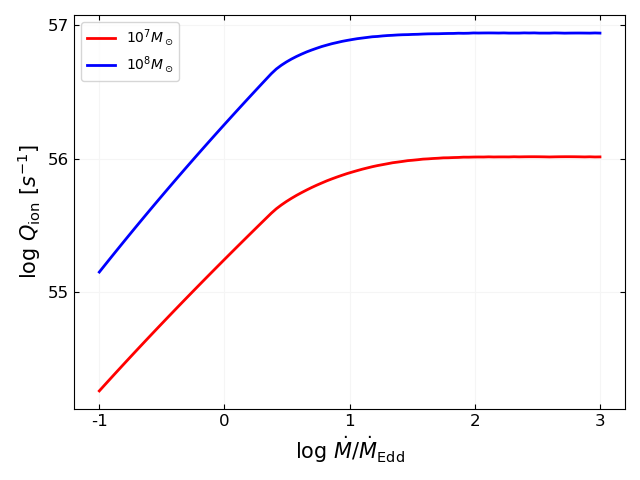}
    \caption{The saturation of \Q for high values of $\Mdot/\dot{M}_{\rm Edd}$ for spinless black holes with masses of $10^7\,\Msun$ (red) and $10^8\,\Msun$ (blue). All the calculations presented here use the slim disk model.}
\label{fig:Q_saturation}
\end{figure}

This drop can be explained by two effects. First, there is a ``saturation'' in \Q at high Eddington ratios due to a saturation in the luminosity of the slim disk, which is driven by the advection of photons from the optically thick regions of the accretion flow onto the BH (horizon), before they are able to escape the flow. We note that this photon advection (or ``trapping''; including in outflows) is a rather generic feature of super-Eddington accretion models, and is not specific to the model we employ here (see discussion in \citealt{KD19} and also, e.g., \citealt{Ohsuga2005,DotanShaviv2011,McKinney14,SN16}, and references therein). This saturation in \Q is demonstrated in Figure \ref{fig:Q_saturation}, which presents \Q as a function of \MMedd for non-spinning black holes with masses of $10^7\,\Msun$ (red) and $10^8 \,\Msun$ (blue), for the slim-disk model. As can be seen in Figure~\ref{fig:Q_saturation}, beyond $\Mdot\simeq{\rm few}\times\dot{M}_{\rm Edd}$ the accretion rate can increase while the ionizing output remains (roughly) constant.

The second effect contributing to the sharp change in \Q is driven by the decrease in BH mass. As was shown in the top panel of Figure \ref{fig:diff_SEDs}, decreasing the BH mass beyond a certain value does not increase the ionizing output of the disk, and may even decrease it. 
As mentioned in Section~\ref{sec:methods_single_BH}, this occurs since for lower BH masses and higher disk temperatures the average energy of the (ionizing) photons increases, and thus the {\it total} number of ionizing photons {\it decreases} (for a given $\Lbol\propto\Mdot$).
We performed a simplified calculation, presented in Appendix \ref{Appendix2}, that suggests the two effects---of photon advection and of decreasing BH mass---can be of comparable significance to the drop in ionizing output in the low-$M$, high-$\MMedd$ regime.

We finally note that there are other models of super-Eddington accretion flows onto SMBHs that, contrary to \texttt{AGNslim}, suggest a {\it suppression}, not enhancement, of UV radiation, driven by the dominance of advection in the inner parts of the flow \cite[e.g.,][]{Ohsuga2005,Pognan20}. 
Such a scenario would naturally lead to yet lower ionizing outputs, and thus yet sharper drops in \Q as SMBHs exceed their Eddington limit, further limiting the period during which AGNs could have contributed to (late-stage) reionization.

% comment
To conclude, our calculations show that slim accretion disks produce (roughly) similar ionizing outputs to those of standard, Eddington-limited thin disks, even when taking into account the excess UV emission produced by the additional Comptonization in the slim disk model we used. 
However, slim disks {\it can} show drastic changes of orders of magnitude in their ionizing output during periods of significantly super-Eddington accretion periods, on relatively short time scales.

\subsection{Extending the contribution of AGNs to reionization to earlier times}
\label{sec:discussion_extending}

In Figure \ref{fig:Nion_diff_spin_fEdd} we demonstrated that a slowly growing population of SMBHs, due to low \fedd and/or high spin, will produce a high number of ionizing photons at high redshifts, as compared to a faster-evolving population (which may be even our fiducial case with $\fedd=0.6$ and $a=0$). However, such slow mass growth scenarios, would inventively lead to require that BHs have high initial masses, that is, massive BH seeds. 

The seed masses at $z=20$ required to produce a $10^9\,\Msun$ SMBH at $z=6$ for each of the scenarios shown in Figure~\ref{fig:Nion_diff_spin_fEdd} are listed in Table \ref{tab:BH_seeds}. 
Specifically, a scenario where all SMBHs have a constant $\fedd=0.1$ and/or a constant maximal spin would imply BH seeds with unreasonably high masses, exceeding $10^7\,\Msun$ (or even $10^8\,\Msun$ if \fedd\ is low {\it and} the spin is high). Such extreme masses exceed even the most massive seeds considered in most works \cite[see reviews by][but see also \citealt{Mayer15} and references therein.]{Natarajan2011,Volonteri12,Inayoshi20}. 
On the other hand, according to our calculations all the growth scenarios shown in Figure \ref{fig:Nion_diff_spin_fEdd} that imply more reasonable BH seed masses ($M\lesssim{\rm few}\times10^4\,\Msun$) would produce an ionizing output that---even by $z=7$---is at least an order of magnitude lower than what is required to keep the IGM ionized (c.f. the gray-shaded area in Fig.~
\ref{fig:Nion_diff_spin_fEdd}.)

\begin{deluxetable}{ccc}
\tablenum{2}
\label{tab:BH_seeds}
\tablewidth{1.0\columnwidth} 
\tablecaption{BH seed masses for the scenarios depicted in Fig.~\ref{fig:Nion_diff_spin_fEdd}.}
\tablehead{
\colhead{$L/L_{\mathrm{Edd}}$} & \colhead{$a$} & \colhead{$M_{\mathrm{seed}}/\Msun$} \\
\colhead{} & \colhead{} & \colhead{at $z{=}20$}
}
\startdata
$0.1$ &  $0$      & $4.2\times10^7$ \\
$0.1$ &  $0.7$    & $1.9\times10^8$ \\
$0.1$ &  $0.998$  & $6.6\times10^8$ \\
$0.6$ &  $0$      & $10$ \\
$0.6$ &  $0.7$    & $4.8\times10^4$ \\
$0.6$ &  $0.998$  & $8.7\times10^7$\\
$1.0$ &  $0$      & $10$ \\
$1.0$ &  $0.7$    & $60$ \\
$1.0$ &  $0.998$  & $1.7\times10^7$
\enddata
\tablenotetext{}{\small{We list the seed masses at $z=20$ required to produce a $10^9\,\Msun$ SMBH at $z=6$.}}
\end{deluxetable}

One way for accreting BHs to maintain a high ionizing output at high redshifts while still having reasonable (implied) seed masses at $z=20$ is to have an extremely early rapid relative growth phase --- immediately after seed formation, followed by a prolonged slow relative growth throughout the epoch of reionization (i.e., most of $z\lesssim6$). 
One such scenario, for a single SMBH, was already explored in Figure \ref{fig:Qvst_Mixed}. In that case, the growth starts with a constant $\fedd=1$ at first and then transitions to a growth at a constant \Mdot (i.e., ``Mixed'' growth). 

\begin{figure}
    \centering
    \includegraphics[width=1.0\columnwidth]{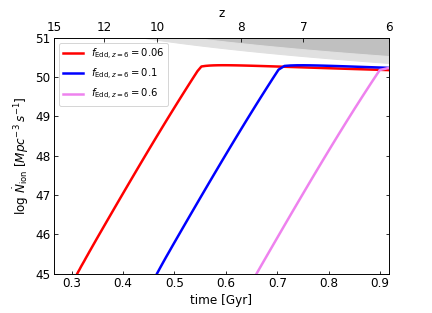}
    \caption{The evolution of \Nion for the population of accreting SMBHs, where the accretion proceeds at an Eddington-limited \Mdot, set by the value of \fedd at $z=6$ (different colors). 
    These calculations assume $a=0$ and ``QLF A'' at $z=6$.}
    \label{fig:Nion_mixed_diff_fEdd}
\end{figure}

This scenario can be applied to the entire SMBH population, described by a QLF, by assuming accretion at a constant spin and some Eddington ratio at $z=6$, as before. 
The results for such a calculation with $a=0$ and various values of $\fedd\left(z=6\right)$ are presented in Figure~\ref{fig:Nion_mixed_diff_fEdd}. 
In these scenarios the ionizing flux of the entire population remains approximately constant during the period of accretion at a constant \Mdot, extending further back in time as the Eddington ratio decreases. 
Specifically, for $\fedd\left(z=6\right)=0.06$, the SMBH population maintains an ionizing flux density higher than $10^{50}\,\Nunits$ up to $z\sim9$. 
Note that for all parameter choices in Figure \ref{fig:Nion_mixed_diff_fEdd}, the implied BH seed masses at $z=20$ are always $\leq10\,\Msun$, while still allowing for the population to include SMBHs with $M=10^9\,\Msun$ by $z=6$ (consistent with the observed $z\simeq6$ quasar population). 

\begin{figure}
    \centering
    \includegraphics[width=1.0\columnwidth]{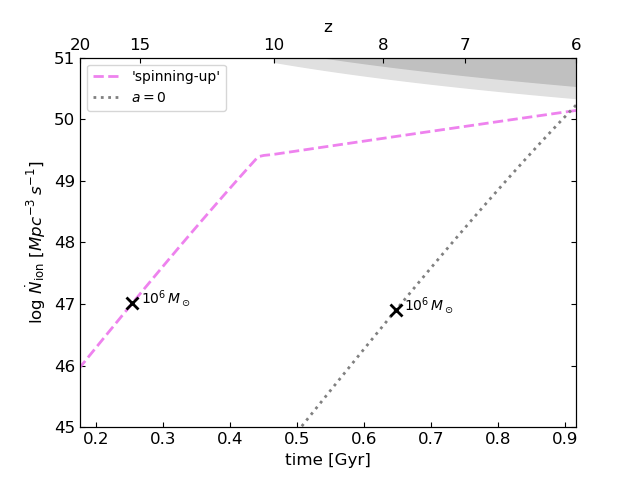}
    \caption{The evolution of the total ionizing flux density of all accreting SMBHs for the simplified ``spin-up'' scenario: $a=0$ for $20<z<10$ and $a=0.998$ for $10<z<6$ (dashed magenta). 
    For comparison, we show again the constant $a=0$ scenario (in dotted gray). 
    For both scenarios we use ``QLF A'' at $z=6$ and assume that the accretion proceeds with a constant $\fedd=0.6$. For each of the scenarios, we also mark the epoch at which a SMBH with $M\left(z=6\right)=10^{9}\,\Msun$ would reach a mass of $10^{6}\,\Msun$.}
    \label{fig:Nion_spin_up}
\end{figure}

Another scenario that achieves a similar effect of high ionizing flux and small BH seeds is that of ``spinning-up'' BHs. In this scenario, accretion begins when the BH has a low spin, followed by the BH being spun-up by the (residual) angular momentum of the disk, until the BH reaches the maximally allowed value ($a=0.998$; see \citealt{King08} and references therein for details of the process). 
We explore a simplified version of such a scenario in Figure~\ref{fig:Nion_spin_up} (dashed magenta), where the spin of the entire BH population is set to $a=0$ for $20<z<10$ ($\Delta t\simeq0.29\,\mathrm{Gyr}$) and then switches to $a=0.998$ for $10<z<6$ ($\Delta t\simeq0.45\,\mathrm{Gyr}$). During the entire growth, the Eddington ratio is fixed to $\fedd=0.6$ for the entire BH population.
This setup allows for a $10^9\,\Msun$ SMBH at $z=6$ to grow from a $z=20$ BH seeds of $\lesssim\,10^6\,\Msun$, which is consistent with the maximal mass proposed for so-called ``direct-collapse'' BH seeding models \citep{Inayoshi20}. 
Compared with the constant high-spin scenario, where the entire BH population has a maximal spin throughout its entire growth history and therefore requires unreasonably high seed masses ($>10^6\,\Msun$ at $z=20$), the ``spin-up'' scenario results in only $10\%$ less ionizing photons in total (i.e., time integrated). 
For comparison, Figure~\ref{fig:Nion_spin_up} also shows the ionizing flux density of a population with $a=0$, and the same \fedd. It can be clearly seen that in such a scenario the drop in ionizing flux with increasing redshift is much more significant. Both scenarios in Figure~\ref{fig:Nion_spin_up} also show the time and redshift at which a $z=6$ $10^9\,\Msun$ SMBH reaches a mass of $10^6\,\Msun$.
Obviously, more complex, physically motivated spin evolution scenarios may be considered  \cite[e.g.,][]{King08,Dotti13}, however these are beyond the scope of the present paper.

We note that in all growth scenarios explored in this work we only considered continuous BH accretion, i.e. a duty cycle of 100\%. 
Any lower duty cycle that results in comparable ``final'' BH masses (at $z=6$) would necessarily extend the accretion to higher redshifts, and thus increase the AGN contribution to reionization over these earlier epochs.
This is similar to the effect of assuming a lower \fedd and/or higher BH spin (i.e., slower mass growth). 
However, we stress that in any Eddington-limited scenario, the duty cycle of the highest mass SMBHs cannot be too low, or else they will not reach their observed high masses at $z\simeq6$.

\subsection{How do accreting SMBHs compare to galaxies?}
\label{sec:discussion_galaxies}

Many studies have tried to identify the sources that have contributed the most to the reionization the Universe, with most of these studies (but not all) concluding that accreting SMBHs are a subdominant source of ionizing radiation as compared to young, star-forming galaxies \cite[e.g.,][and references therein]{Robertson15, Bouwens15, Parsa18, Dayal20, Ananna20}. 
In this section we compare the results of our SMBH-focused calculations to the ionizing contribution of such galaxies, as reported by the comprehensive study by \cite{Bouwens15}. 
That study derived the \Nion attributed to early star forming galaxies, according to
\begin{equation}
    \label{eq:Bouwens Nion}
    \Nion = \fesc\xiion \rho_\mathrm{UV}\,,
\end{equation}
where $\rho_\mathrm{UV}$ is the observed rest-frame UV luminosity density of Lyman-break galaxies ($\mathrm{erg}\,\mathrm{s}^{-1}\,\mathrm{Hz}^{-1}\,\mathrm{Mpc}^{-3}$); 
\xiion is the Lyman-continuum photon production efficiency; 
and \fesc is the fraction of hydrogen-ionizing photons that escape the galaxies and affect the IGM (the escape fraction). 
It then calculated the empirical evolution of \Nion, as constrained by measurements of the Thompson optical depth of CMB photons and various other astrophysical probes of the ionization state of the Universe. Comparing the two ionizing emissivities, \cite{Bouwens15} concluded that there is a good agreement between the observed \Nion and the one derived from the observed galaxy population, in both redshift evolution, as well as in normalization, provided that $\log\fesc \xiion =24.50$ is assumed for all galaxies. 
There are significant uncertainties on both \fesc and \xiion \cite[e.g.,][]{Siana10, Vanzella12, Mostardi13, DC15}, with the \cite{Bouwens15} study considering the ranges  $\log\fesc\simeq(-1.3)-(-0.4)$ and $\log\left(\xiion/\rm{s}^{-1}/\left(\rm{erg}\,\rm{s}^{-1}\,\rm{Hz}^{-1}\right)\right)\simeq25.2-25.5$. 

\begin{figure}
    \centering
    \includegraphics[width=1.0\columnwidth]{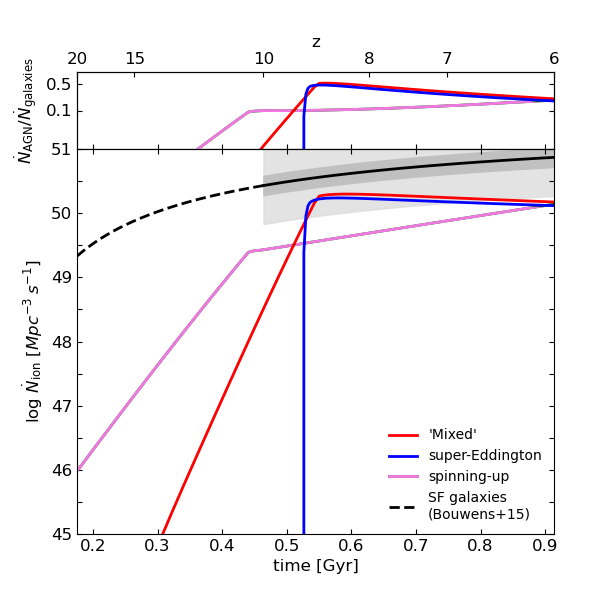}
    \caption{Comparison of \Nion for the most ``optimistic'' AGN-focused scenarios presented in this work: the ``Mixed'' (red) and super-Eddington (blue) scenarios with $\fedd\left(z=6\right)=0.06$ and the ``spin-up'' scenario (magenta). 
    All these scenarios use ``QLF A'' at $z=6$. 
    The \Nion of galaxies, as derived by \cite{Bouwens15} assuming $\log\fesc\xiion=24.50$, is shown in solid black for $6<z<10$, while the dashed line traces its extrapolation over $10<z<20$. Dark- and light-shaded gray regions correspond to the reported range of \xiion and the additional range of \fesc, respectively. The top panel presents the ratio between the \Nion of AGNs and that of galaxies, for the three AGN scenarios.}
    \label{fig:Bouwens15}
\end{figure}

Figure~\ref{fig:Bouwens15} shows the evolving \Nion derived for young, star-forming $6<z<10$ galaxies by \cite{Bouwens15}, under the assumption of a universal $\log\fesc\xiion=24.50$ (solid black line), along with the extrapolation to earlier epochs (dashed black line), as well as the appropriate uncertainty ranges (shaded regions; see caption). 
Figure~\ref{fig:Bouwens15} also shows the most ``optimistic'' scenarios for the population of growing SMBHs explored in the present work, judged by their ionizing emissivity over relatively long periods.
Specifically, we show the ``Mixed'' and super-Eddington scenarios with $\fedd\left(z=6\right)=0.06$ (red and blue lines, respectively) and the simplified ``spin-up'' scenario (magenta). 
For all the scenarios plotted here we assumed the ``QLF A'' at $z=6$.
To ease the comparison between the ionizing output of AGNs (following our calculations) and that of galaxies (following \citealt{Bouwens15}), the top panel in Fig.~\ref{fig:Bouwens15} shows the ratio between the former and the latter (i.e., $\dot{N}_{\mathrm{AGN}}/\dot{N}_{\mathrm{galaxies}}$).

Several conclusions can be drawn from Figure \ref{fig:Bouwens15}. First, provided that $\log\fesc\xiion=24.50$, the ionizing flux density of galaxies is very high at $z=6$, at roughly   (${\sim}7\times10^{50}\,\Nunits$) and remains high up to $z{\simeq}10$ ($\gtrsim 2.7\times10^{50}\,\Nunits$). For comparison, for AGNs in the ``Mixed'' and super-Eddington scenarios, the \Nion is in the range of $(1-2)\times10^{50}\,\Nunits$ between $6<z<9$, which is subdominant to the one found by \cite{Bouwens15} by a factor of few. 
For the ``spin-up'' scenario the AGNs-to-galaxies ratio is slightly lower, as the AGN \Nion drops below $10^{50}\,\Nunits$ before reaching $z=7$. Note, however, that for the lower end of the uncertainty in $\log\fesc\xiion$ the ionizing flux density of AGNs can be comparable to that of galaxies, and can even surpass it at higher redshifts. 

An important caveat here is that throughout our analysis we assumed that all the ionizing radiation of accreting SMBHs escapes to the IGM. In this regard the results presented here serve as an upper limit on the AGNs' \Nion. In addition, the AGN scenarios in Figure \ref{fig:Bouwens15} are somewhat contrived: the ``spin-up'' scenario requires very massive seeds (see Section~\ref{sec:res_population}), while the ``Mixed'' and super-Eddington scenarios require a very low \fedd of $0.06$ at $z=6$, in tension with observations of the most luminous quasars. To demonstrate the latter point, we note that \cite{Shen19} found an average $\fedd\simeq0.3$ for a large sample of $z\gtrsim5.7$ quasars, and that \cite{Onoue19} found $0.16\lesssim\fedd\lesssim1.1$ for a sample of six fainter quasars at $z>5.8$ \cite[see also][]{Mazzucchelli17}. It is not impossible, however, that the more typical \fedd of the entire $z=6$ AGN population, including the yet-to-be-detected low-luminosity AGNs \cite[e.g.,][]{Weigel15,Cappelluti16,Vito2016}, is indeed closer to the $\fedd = 0.06$ assumed in our ``optimistic'' scenarios.\footnote{See, e.g., \cite{Schulze2015} for an example of how the ``typical'' \fedd is determined for complete AGN samples at lower redshifts.}

Thus, we conclude that it is unlikely that accreting SMBHs were the main drivers of reionization at high redshifts. However, significant contribution at $z\lesssim 8$ is plausible, and can even be dominant at higher redshifts, provided that (1) galaxies have a relatively low $\log\fesc\xiion$, and (2) SMBHs grow through certain scenarios while the escape fraction for the emergent AGN radiation remains high.

\subsection{Ionized regions around quasars}
\label{sec:discussion_ionized_regions}

To demonstrate the applicability of the framework developed in this work, we use the ionizing fluxes derived in several growth scenarios to calculate the sizes of `quasar near zones', which are H\,{\sc ii} regions around quasars with sizes of order ${\sim}0.1-10\,\mathrm{Mpc}$. These ionized regions can be probed by rather direct (spectroscopic) observations, with their sizes potentially probing the quasar activity timescales \cite[e.g.,][]{Eilers17, Eilers20, Davies20, Chen21}. 
In addition, high IGM neutral fractions leave an imprint in the quasar spectra in the form of a Ly$\alpha$ ``damping-wing'', the detection of which may thus probe the high neutral fraction regime, at $z\gtrsim7$ \cite[e.g.,][]{Banados18, Wang20}.

The calculations presented here are simplistic and serve both to demonstrate the implications of our framework for the size evolution of large-scale H\,{\sc ii} regions, as well as provide an independent test for the relevance of our calculations. 

The near zones calculations presented here are based on the method described in detail by \cite{Davies16}. 
In brief, we solve one-dimensional radiative transfer ordinary differential equations, assuming a central ionizing source embedded in pure-hydrogen, fully neutral, uniform- and constant-density IGM, in order to calculate the ionized region size as a function of time.
We have experimented with several slightly more elaborate assumptions (e.g., various choices for the IGM density and ionization fraction), but these did not significantly affect the main results and trends we discuss below.

\begin{figure}
    \centering
    \includegraphics[width=1\columnwidth]{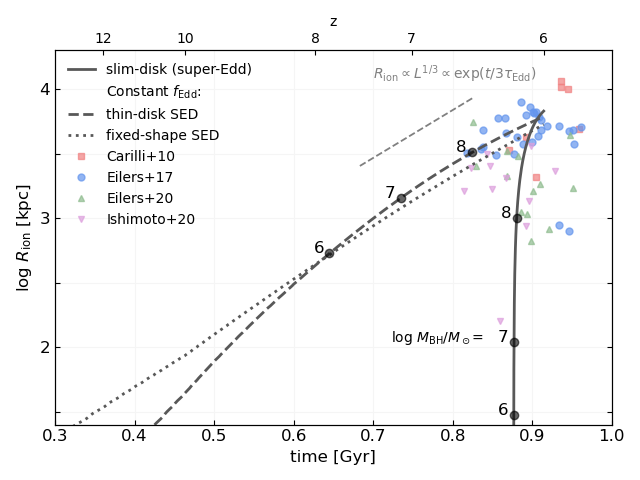}
    \caption{The evolution of $R_\mathrm{ion}$ around a spinless SMBH with $M=10^9\,\Msun$ and $\fedd=0.6$ at $z=6$, for different SED models: slim-disk (solid), thin-disk (dashed) and fixed-shape (dotted). The slim-disk model evolves with a constant \Mdot while the thin-disk and fixed-shape models evolve with a constant \fedd. 
    The different colored symbols mark measurements of luminosity-corrected proximity-zone sizes from \cite{Carilli10}, \cite{Eilers17}, \cite{Eilers20} and \cite{Ishimoto20} (see legend). 
    The dashed gray line shows the relationship $R_{\rm{ion}}\propto\Lbol^{1/3}\propto\exp\left(t/3\tau_{\rm Edd}\right)$, as expected for a fixed-shape SED. The points at which $\log M_{\rm{BH}}/ \Msun$ reaches various values are marked for the slim and thin-disk scenarios.}
    \label{fig:Rion}
\end{figure}

Figure \ref{fig:Rion} presents the evolution of the proper size of the quasar near-zone, $\Rion$, for a spinless SMBH with $M=10^9\,\Msun$ and $\fedd=0.6$ at $z=6$, along with luminosity-corrected measurements of proximity zones around quasars at $6\lesssim z\lesssim7$ taken from \cite{Carilli10}, \cite{Eilers17}, \cite{Eilers20} and \cite{Ishimoto20}. 
Here we define \Rion as the distance from the ionizing source to the point where the hydrogen ionization fraction reaches a value of $0.9$.\footnote{In practice, the ionization state transition is sufficiently sharp so that any analysis is indifferent to this specific choice.}
We show the results of calculations that assume either the thin-disk (dashed line) and fixed-shape (dotted line) SED models, where the accretion proceeds with a constant \fedd, as well as the slim disk model which proceeds at a constant \Mdot, including in the super-Eddington regime (solid line). 

Inspecting Figure~\ref{fig:Rion}, we first note that our calculated $z\simeq6$ proximity zone sizes are reassuringly in general agreement with observations.
The most striking difference between the super-Eddington, slim-disk model and the two Eddington-limited models is that the size evolution of $\Rion$ for the former is much steeper, going from ${\sim}10\,\mathrm{kpc}$ to ${\sim}10^3\,\mathrm{kpc}$ within less than 40 Myr---a factor of ${\sim}15$ faster than for the standard thin-disk, Eddington-limited model (which, in turn, is ${\sim}15\%$ faster as compared to the fixed-shape SED model).
The evolution of the proximity zone is thus much more recent for the super-Eddington model, occurring mostly at $z\lesssim6.5$. In addition, the final proximity zone size is the largest for the slim-disk model, followed by the thin-disk (smaller by ${\sim}10\%$) and then by the fixed-shape SED (smaller by ${\sim}20\%$).

The slightly faster growth of the H\,{\sc ii} region for the standard thin disk than for the constant shape SED is consistent with the sharper increase in \Q for the former (e.g., Fig. \ref{fig:Qvst_const_fEdd}), with the larger ``final'' \Rion being due to the overall higher number of ionizing photons emitted (see Section~\ref{sec:res_single_bh}). 
The much faster growth of the H\,{\sc ii} region for the slim disk model stems from the fast \Q evolution for accretion at super-Eddington rates (Fig. \ref{fig:Qvst_AGNslim}), with the final region size being the largest of the three models due to the hard spectrum of the slim disk SED (Fig. \ref{fig:SED_AGNslim_vs_thin}).

One of the most striking differences between the various scenarios explored in Figure \ref{fig:Rion} is the relation between \Rion and the BH mass $M$. The points at which $\log M_{\rm BH}/ \Msun$ reaches 6,7 and 8 are marked on the curves of the thin and slim-disk scenarios. 
Evidently, although the final $z=6$ \Rion size (i.e., when the BH reaches a mass of $10^9\,\Msun$) is similar for the two accretion scenarios, for any lower BH mass (i.e. for any epoch at $z>6$) the proximity zones for the slim-disk, super-Eddington scenario are much smaller than for the thin-disk, Eddington-limited one. 
For example, in the thin-disk case the proximity zone reaches a size of $1\,{\rm Mpc}$ when the BH has reached only $M\simeq4\times10^6\,\Msun$, while for the slim-disk case the same \Rion is reached only when the BH mass is as massive as $\sim10^8\,\Msun$.

Our results echo, and provide further support for, the idea that $z\simeq6$ quasars with relatively compact proximity zones are ``young'', as explored in several recent works \cite[e.g.,][]{Andika20, Eilers20,Eilers2021,Morey2021}. 
Within our framework and the calculations presented in Figure~\ref{fig:Rion}, the only way to obtain small proximity zones ($\lesssim$1 Mpc) around $z\simeq6$ quasars that are powered by ``mature'' SMBHs with $M\gtrsim10^8\,\Msun$, is if their growth has started at an extremely late stage ($z<7$) and proceeded through extremely fast, super-Eddington accretion, to be able to reach the ultimate high BH masses.
Unless the escape fraction is extremely low (i.e., heavily obscured systems), the key for producing compact proximity zones is a low (mass-averaged) radiative efficiency of the accretion process, as is the case for super-Eddington accretion.
Alternatively, a compact proximity zone could result from low duty cycle BH activity (i.e., sub-Eddington accretion turning on and off intermittently), however the resulting ``final'' BH mass may be significantly lower than what is observed for the $z=6$ quasar population.
At any rate, the calculations highlighted in this Section demonstrate that our framework can be used as a predictor of (trends in) $R_\mathrm{ion}$, and as an efficient way to link various BH growth scenarios to proximity zone observations.

%%%%%%%%%%%%%%%%%%%%%%%%%%%%%%%%%%%%%%%%%%%%%%%%%%%%%%%%%%%%%%%%
\section{Summary and Conclusions}
\label{sec:summary}

Any attempt to assess the contribution of accreting SMBHs to the reionization of the Universe has to consider several (sometimes contradictory) aspects: a SMBH produces ionizing radiation only during vigorous accretion episodes; the faster it grows --- the more ionizing radiation it produces; grow \textit{too} fast, and the contribution to reionization is limited to a short period; grow too \textit{slow}, and the high observed BH masses cannot be explained. 
To complicate things further, additional (subtle) effects are expected when considering a population of growing SMBHs, and their spectral evolution.

In this work we presented a framework for the calculation of the ionizing output of accreting SMBHs in a physically motivated way that accounts for the growth of the SMBH and the spectral dependence on BH mass, accretion rate, and spin. We then extended the framework to a population of SMBHs, assuming a QLF representation and a universal \fedd (defined as $L_\mathrm{bol}/L_\mathrm{Edd}$) at $z=6$. After modeling the mass and spectral evolution of the population, we derived the history of the ionizing flux density of the entire SMBH population.

The key results from the application of this framework are:
\begin{itemize}
    \item Accounting for the spectral evolution of accreting SMBHs can increase their total ionizing output by $\sim30-80\%$, compared to a (commonly-used) fixed-shape SED model (Figs. \ref{fig:Qvst_const_fEdd}, \ref{fig:Qvst_Mixed}, \ref{fig:Qvsz}).
    
    \item Accreting SMBHs can probably contribute significantly to cosmic (hydrogen) reionization only at late times ($z\lesssim7$; Figs. \ref{fig:Qvsz}, \ref{fig:Nion_diff_spin_fEdd}). 
    
    \item Slower mass growth of the SMBH population, by means of a low \fedd and/or high spin, increases significantly the population's ionizing output at high redshifts ($z>7$), and may allow for a non-negligible contribution of accreting SMBHs to reionization at up to $z\approx9$ (Figs. \ref{fig:Nion_diff_spin_fEdd}, \ref{fig:Nion_mixed_diff_fEdd}, \ref{fig:Nion_spin_up}).
    
    \item Growth scenarios with periods of super-Eddington accretion only highlight the previous points: a late contribution to reionization, which can be extended to higher redshifts by means of a lower accretion rate, but with a yet sharper drop in the ionizing output at earlier epochs (Fig. \ref{fig:Nion AGNslim}).
    
    \item Accounting for the spectral evolution of accreting SMBHs can slightly increase the size of H\,{\sc ii} regions around quasars (by ${\sim}10\%$). The super-Eddington slim disk model can increase the size by a further ${\sim}10\%$, and --- more importantly --- lead to a very fast growth (a factor of $\times15$ faster than for the standard thin disk; Fig. \ref{fig:Rion}).

    \item SMBHs are probably a sub-dominant source of ionizing radiation, as compared to galaxies. However, the relative contribution of SMBHs to reionization can be increased by a slow mass evolution at low redshifts ($6\lesssim z \lesssim9$); and/or an exceptionally high space density of moderate-luminosity AGNs; and/or if the ionizing radiation of star-forming galaxies is suppressed (intrinsically and/or by a low escape fraction; see Fig. \ref{fig:Bouwens15}).
    
\end{itemize}
The last point notwithstanding, it is important to note that the recent measurement of a low Thompson optical depth of CMB photons made by \cite{Planck18VI} is consistent with a late ($z<9$) reionization, with \cite{Mason19}, for example, finding $z_{0.5}=6.93\pm0.14$ as the redshift of the mid-point of reionization. 
Moreover, other recent studies, which are based on observations (Ly$\alpha$ forest  measurements) or models (radiative transfer or hydrodynamical calculations), have concluded that reionization may have extended up to $z\sim5.3$ \cite[e.g.,][]{Eilers18,Kulkarni19,Keating20,Bosman18,Bosman21,Zhu21}. When taken together with the results of this work, namely the late contribution of AGNs to reionization and possible extension of the AGN contribution towards $z\sim9$, this makes the scenario in which AGN contribution to reionization is non-negligible (and even comparable to that of galaxies) more plausible.

We note that all of our \Q and \Nion calculations can be trivially shifted to earlier times (e.g., setting the end-point of the BH evolution at some $z>6$) with the only caveat being higher implied seed masses.

There is a great degree of uncertainty in the population analysis presented in Section~\ref{sec:res_population} due to uncertainties involving the QLF, specifically at $z\gtrsim6$. Currently there is very limited knowledge of the lower-luminosity shape of the QLF due to a lack of observations of relevant AGNs beyond $z\simeq4.5$ \cite[e.g., ][]{Shen20}.

Other uncertainties are related to the simplifying assumptions we made throughout the present work, each of which could be the focus of future investigations. One such assumption is that of a constant spin, in contrast to a scenario of a BH with a self-consistent spin evolution \cite[e.g.,][]{King08,Dotti13,Volonteri13}. In addition, when interpreting the QLF, we assumed a fixed \fedd and spin for the entire population, while the actual population may have a wide range of Eddington ratios \cite[e.g.,][]{Mazzucchelli17,Shen19} and spins, and also assumed that new BHs are not being formed, nor that AGNs turn their accretion  ``on'' or ``off''. 
Furthermore, a more detailed analysis of the radiative outputs of SMBHs should consider the total number of ionizations in the IGM, including secondary ones, in contrast to the simpler approach taken here (focusing only on \Q and \Nion). 
This may be particularly relevant for quasars, due to their hard SEDs, and even more relevant for slim-disk, super Eddington models (such as \texttt{AGNslim}), where the SEDs are even harder.

In the present work, we have not considered the physics of circumnuclear, interstellar, and/or intergalactic media, which would include obscuration, attenuation and gas geometry. Perhaps the best way to address these complex processes is by incorporating the SEDs and considerations that were highlighted in this work as ``sub-grid'' components in large cosmological  hydrodynamic simulations. 
Such simulations, and/or semi-analytical models, can also be used to explore various other BH formation and early growth scenarios, which are not captured by the QLF-based population analysis presented here.
Going beyond hydrogen ionization, our framework could be also extended to investigate the contribution of accreting SMBHs to (later) helium reionization and to early cosmic heating.

The detection of ever larger and more complete populations of early accreting SMBHs, beyond $z\simeq7$, coupled with advances in the understanding of their accretion flows and growth histories, should lead to further re-assessment of the contribution of SMBHs to the reionization of the Universe.

%%%%%%%%%%%%%%%%%%%%%%%%%%%%%%%%%%%%%%%%%%%%%%%%%%%%%%%%%%%%%%%%
% \acknowledgments

\begin{acknowledgments}

We thank the anonymous reviewer for their constructive and insightful comments, which helped us improve this paper.
We thank Smadar Naoz for early discussions that motivated key parts of our work, and Rennan Barkana for useful comments that helped improve this paper.
We thank Aya Kubota and Chris Done for their assistance in incorporating the slim-disk model into our framework and for helpful comments.
We also thank Shane Davis for his assistance with some of the \texttt{KERRTRNAS} calculations. 
We finally thank Steven Furlanetto for kindly providing the energy deposition fractions for the radiative transfer calculations.
We acknowledge support from the Israel Science Foundation (grant number 1849/19) and from the European Research Council (ERC) under the European Union's Horizon 2020 research and innovation program (grant agreement number 950533).

\end{acknowledgments}

\software{{\tt AstroPy} \citep{Astropy2013,Astropy2018},
{\tt Matplotlib} \citep{Matplotlib2007}, 
{\tt NumPy} \citep{NumPy20}, {\tt SciPy} \citep{SciPy20}, {\tt Xspec and \tt PyXspec} \citep{Xspec96}}

% \smallskip
\clearpage
\bibliography{SMBHs_disks_reion}{}
\bibliographystyle{aasjournal}

\appendix
\section{Testing the angle dependence of disk SEDs} \label{Appendix0}

In Section~\ref{sec:methods_single_BH} we mentioned our choice to ignore the angle dependence of thin-disk SEDs calculated for (non-spinning) SMBHs.

\begin{figure}[h!]
    \centering
    \includegraphics[width=0.475\textwidth]{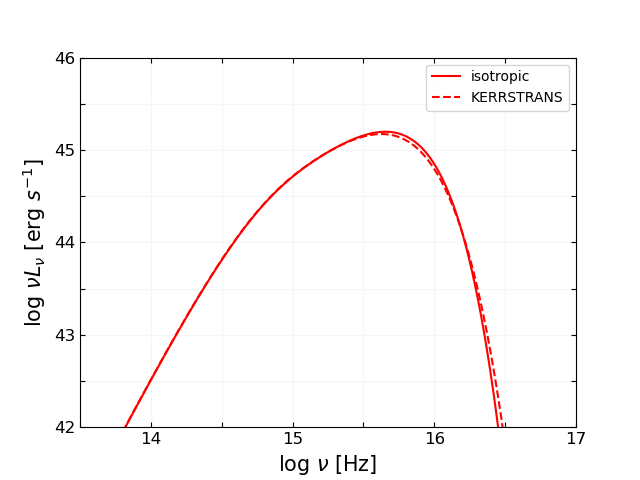}
    \caption{A comparison of a thin-disk SED calculated either by ignoring non-isotropic relativistic effects (solid) or taking them into account (using \texttt{KERRTRANS}; dashed).
    Both cases assume a non-spinning BH with $M=10^8\,\Msun$ and $\Mdot=1\,\mpyr$.}
    \label{fig:kerrtrans vs regular}
\end{figure}

To demonstrate the validity of this choice, in Figure~\ref{fig:kerrtrans vs regular} we show thin-disk SEDs calculated with and without taking into account the angle dependence. 
The solids line shows an SED calculated while ignoring non-isotropic relativistic effects, while the dashed line shows an SED calculated based on a series of inclination-dependent SEDs, each of which is produced using \texttt{KERRTRANS}, and then integrating them over all lines-of-sight.
In both cases, the SEDs assume a non-spinning BH with $M=10^8\,\Msun$ and $\Mdot=1\,\mpyr$. For these parameters the ``isotropic'' SED (solid) results in $8\%$ higher production rate of ionizing photons than the \texttt{KERRTRANS} SED.

% \label{Appendix1}

\section{Demonstrating thin-disk SED evolution} \label{Appendix1}

\begin{figure}[h!]
    \centering
    \includegraphics[width=0.475\textwidth]{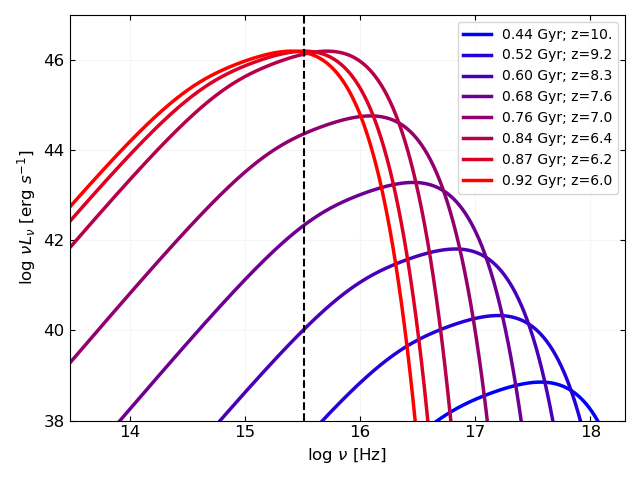}
    \caption{Standard thin-disk SEDs at different times and redshifts for the ``Mixed'' growth scheme with $\Mdot = 10\,M_\odot/\rm{yr}$ (blue curve in Fig. \ref{fig:Qvst_Mixed}). The Lyman limit is marked by a vertical dashed line.}
    % caption
    \label{fig:SED_evolution_demo}
\end{figure}

The evolution of the SED of a thin accretion disk SED for a SMBH accreting at an Eddington-limited, constant $\Mdot=10\,\Msun/\rm{yr}$ (Fig. \ref{fig:Qvst_Mixed}; blue curve) is presented in Figure \ref{fig:SED_evolution_demo}. 
At late times, the SEDs move horizontally due to the increasing BH mass, causing a decrease in \Q, while in earlier time, during the Eddington-limited accretion growth phase, the SEDs move both horizontally and vertically, causing an exponential increase in \Q with time.

\section{The saturation in \Q for low-$M$ and high-$\MMedd$} \label{Appendix2}

\begin{figure}
\centering
    \includegraphics[width=0.475\textwidth]{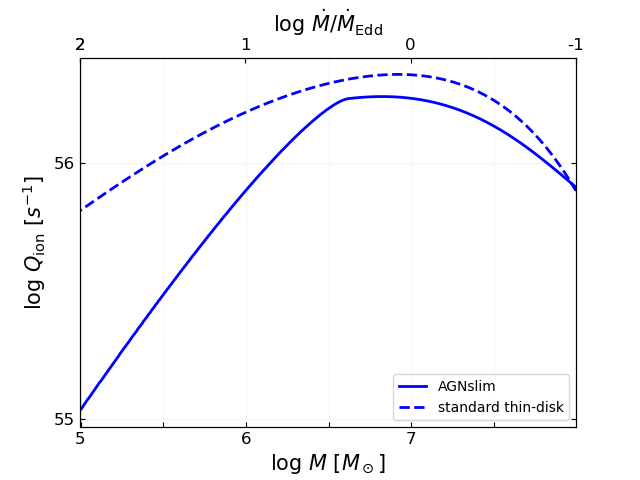}
    \caption{\Q as a function of $M$, and the corresponding $\Mdot/\dot{M}_{\rm Edd}$ (top axis), for the slim disk model (solid) and the standard thin-disk model (dashed), assuming a constant  $\Mdot=4.5\,\mpyr$ and $a=0$. 
    Both models exhibit a drop in \Q for low masses, as the same overall radiative output (Eq.~\ref{eq:Lbol}) is carried by higher-energy photons. The slim disk model has an additional drop in \Q in early epochs (i.e., highest accretion rates), due to saturation (see text for details).
    For the purposes of this plot, the thin-disk model is extrapolated beyond the Eddington limit to illustrate the relative significance of the saturation.}
\label{fig:saturation_mass}
\end{figure}

In Section~\ref{sec:discussion_superEdd} we have discussed the two effects contributing to the ``saturation'' in \Q in the low-$M$, high-$\MMedd$ regime.
We have demonstrated the saturation in \Q for super-Eddington accretion in Figure \ref{fig:Q_saturation}, which presents \Q as a function of \MMedd. 

Another demonstration of the two effects driving the sharp change in \Q is presented in Figure \ref{fig:saturation_mass}. 
Here, \Q is plotted as a function of $M$ (and \MMedd in the top horizontal axis), for a constant value of $\Mdot=4.5\,\mpyr$ and $a=0$. 
For a standard thin disk around a non-spinning BH, this \Mdot\ would corresponds to $\fedd=0.1$ for a BH with $M=10^9\,\Msun$.
The solid line is obtained from the slim disk model and the dashed line from the standard thin disk. Note that the standard thin disk model is not valid in the super-Eddington regime, and is plotted here only to demonstrate the decrease in \Q caused {\it solely} by the increasing disk peak temperature associated with a decreasing BH mass, which is generally expected for accretion disks around SMBHs \citep{LyndenBell69}.

For the slim disk model, which is affected by photon advection (and thus further ``saturation''), \Q decreases by a factor of $\gtrsim15$ between the peak in ionizing output emission and $\MMedd\simeq100$.
In comparison, the corresponding drop for the thin-disk model (extrapolated beyond its realistic domain) is only a factor of $\sim(3{-}4)$.
Thus, the advection of photons from the inner parts of the slim disk suppresses the ionizing photon flux significantly, regardless of the effect of decreasing BH mass.

\end{document}

% End of file `sample63.tex'.